\documentstyle[11pt,aaspp4,flushrt,tighten,psfig]{article}  
\newcommand{\mincir}{\raise
  -2.truept\hbox{\rlap{\hbox{$\sim$}}\raise5.truept \hbox{$<$}\ }}
\newcommand{\magcir}{\raise
  -2.truept\hbox{\rlap{\hbox{$\sim$}}\raise5.truept \hbox{$>$}\ }}

\begin{document}

\title{Iron abundance in the ICM at high redshift}

\author{P. Tozzi\altaffilmark{1}, P. Rosati\altaffilmark{2},
S. Ettori\altaffilmark{2}, S. Borgani\altaffilmark{3},
V. Mainieri\altaffilmark{2,4}, C. Norman\altaffilmark{5}}

\affil{$^1$INAF, Osservatorio Astronomico di Trieste, via G.B. Tiepolo
11, I--34131, Trieste, Italy} \affil{$^2$European Southern
Observatory, Karl-Schwarzschild-Strasse 2, D-85748 Garching, Germany}
\affil{$^3$Dip. di Astronomia dell'Universit\`a, via G.B. Tiepolo 11,
I--34131, Trieste, Italy}
\affil{$^4$Dip. di Fisica, Universit\`a degli Studi di Roma III, via
della Vasca Navale 84, I--00146 Roma, Italy}
\affil{$^5$ Department of Physics and Astronomy, Johns Hopkins University,
Baltimore, MD 21218}

\begin{abstract}

We present the analysis of the X--ray spectra of 18 distant clusters
of galaxies with redshift $0.3 < z < 1.3$.  Most of them were observed
with the Chandra satellite in long exposures ranging from 36 ks to 180
ks.  For two of the $z>1$ clusters we also use deep XMM--Newton
observations.  Overall, these clusters probe the temperature range $3
\, \mincir\, kT\, \mincir \, 8$ keV.  Our analysis is aimed at
deriving the iron abundance in the Intra Cluster Medium (ICM) out to
the highest redshifts probed to date.  Using a combined spectral fit
of cluster subsamples in different redshift bins, we investigate the
evolution of the mean ICM metallicity with cosmic epoch. We find that
the mean Fe abundance at $\langle z \rangle =0.8$ is $Z =
0.25^{+0.04}_{-0.06} \, Z_\odot$, consistent with the local canonical
metallicity value, $Z\simeq 0.3 \, Z_\odot$, within 1$\sigma$
confidence level.  Medium and low temperature clusters ($kT <5 $ keV)
tend to have larger iron abundances than hot clusters.  At redshift
$\langle z\rangle \sim 1.2$ (4 clusters at $z>1$) we obtain a
statistically significant detection of the Fe-K line only in one
cluster ($Z>0.10 Z_\odot$ at the 90\% c.l.).  Combining all the
current data set from Chandra and XMM at $z>1$, the average
metallicity is measured to be $\langle Z\rangle =
0.21^{+0.10}_{-0.05}\, Z_\odot$ (1$\sigma$ error), thus suggesting no
evolution of the mean iron abundance out to $z\simeq 1.2$.

\end{abstract}


\newpage

\section{INTRODUCTION}

Spectroscopic observations in the X--ray band provide a powerful means
to probe the metal content of the diffuse gas in groups and clusters
of galaxies.  Typical metallicity in the Intra--Cluster Medium (ICM)
of rich clusters out to redshifts $z\mincir 0.4$, are close to the
{\sl canonical} value of $Z = 0.3 \, Z_\odot$ (Mushotzky \&
Loewenstein 1997; Fukazawa et al. 1998; Allen \& Fabian 1998; Della
Ceca et al. 2000; Ettori, Allen \& Fabian 2001), whereas they are
found to be somewhat higher (up to a factor of two) in moderate mass
clusters and groups (Buote 2000; Finoguenov, Arnaud \& David 2001;
Davis, Mulchaey \& Mushotzky 1999; but see Renzini 1997).  These
observations show clearly that a significant amount of the metals
produced by supernovae are injected into the ICM, a process which is
essentially completed by $z\sim 0.4$.

The presence of metals in the ICM is crucial for tracing the past
history of star formation in cluster galaxies. Supernova explosions,
which are the main contributor to metal enrichment, may also provide a
significant source of heating for the ICM.  A number of observations
concerning the ICM X--ray scaling properties (see Ponman, Cannon \&
Navarro 1999; Rosati, Borgani \& Norman 2002 and references therein;
Vikhlinin et al. 2002) suggests that non--gravitational heating
processes must have taken place in the past and significantly affected
the thermodynamics of the diffuse gas (e.g.  Tozzi \& Norman 2001;
Bialek, Evrard \& Mohr 2001; Babul et al. 2002; Tornatore et al. 2003;
Voit et al. 2002).  Therefore, a knowledge of the metal content of
galaxy clusters and its evolution, combined with a precise modeling
of the star formation processes, allows one to estimate the total
amount of energy supplied by supernova explosions (e.g., Pipino et
al. 2002, and references therein) and, ultimately, to understand the
interplay between the evolution of the hot diffuse baryons and of the
cold gas locked in the stellar phase.

On the observational side, data from the ASCA and Beppo-SAX satellites
have allowed us to trace the spatial distribution of metals for a
fairly large number of clusters (e.g., White 2000; Finoguenov, David
\& Ponman 2000; Dupke \& White 2000; De Grandi \& Molendi 2001;
Finoguenov et al. 2002.)  Such studies have provided useful
information on the connection between the ICM metal distribution, the
dynamical status of the clusters and the stellar population which
synthesized such metals.  For instance, De Grandi \& Molendi (2001)
found evidence for Fe gradients in relaxed systems which show
signatures of cooling cores (but see Lewis, Buote \& Stocke 2003). By
mapping the different distributions of Ne, Si, S and Fe, Finoguenov et
al. (2000) argued that Type Ia SN provide a larger contribution than
Type II SN to the enrichment of central regions of clusters.  With the
advent of the Chandra and XMM satellites, and their unprecedented
sensitivity and angular resolution, these studies are now carried out
in great detail (e.g., Tamura et al. 2001; Ettori et al. 2002;
Gastaldello \& Molendi 2002; Matsushita et al. 2002), and provide new
strong constraints on the physics of the ICM, as well as its chemical
properties.

Specifically, with Chandra and XMM deep pointings on distant clusters,
one can investigate the evolution of chemical abundances beyond
$z=0.5$, thus extending a previous analysis based on ASCA data, which
has shown no significant evolution in the iron abundance out to $z\sim
0.4$ (Mushotzky and Loewenstein 1997; Matsumoto et al. 2000).  By
tracing the evolution of the global metal content of the ICM one
obtains useful information on the epoch at which the last significant
episode of star formation in cluster galaxies took place and enriched
the diffuse baryons.

In this Paper, we measure iron abundances in the ICM from a sample of
18 distant clusters extracted from the Chandra and XMM archives. Most
of the analysis is based on Chandra data (14 pointings for 17
targets); two additional XMM observations are used for two clusters at
$z>1$, one of which (RXJ0849) was also targeted by Chandra.  The whole
sample covers the redshift range $0.3\mincir z \mincir 1.3$ and
includes clusters with temperature between 3 and 8 keV.

The plan of the Paper is as follows. In \S 2 we describe the data
reduction procedure. In \S 3 we describe the spectral analysis of the
single sources (\S 3.2), and of the combined spectra of subsamples (\S
3.3). In \S 4 we discuss possible implications of our findings and
summarize our conclusions.  We adopt a cosmological model with $H_0 =
70$ km/s/Mpc, $\Omega_M=0.3$ and $\Omega_\Lambda = 0.7$ throughout.
Quoted confidence intervals are 68\% unless otherwise stated.

\section{OBSERVATIONS AND DATA REDUCTION}

\subsection{Chandra data}

In Table \ref{exposures}, we present the list of the Chandra
observations analyzed in this Paper.  Some of them are also part of
the analysis presented by Ettori, Tozzi \& Rosati (2003).  Most the
observations were carried out with ACIS--I, while for MS2137, MS1054
and MS0451 the Back Illuminated S3 chip of ACIS--S was also used.  The
data are reduced using the 2.3 version of the CIAO software (release
V2.3, see {\tt http://cxc.harvard.edu/ciao/}), and use the level=1
event file as a starting point.  For observations taken in the VFAINT
mode, we run the tool {\tt acis\_process\_events} to flag probable
background events using all the information of the pulse heights in a
$5\!\times \!5$ event island (as opposed to a $3\!\times \!3$ event
island recorded in the FAINT mode) to help distinguishing between good
X--ray events and bad events that are most likely associated with
cosmic rays.  With this procedure, the ACIS particle background can be
reduced significantly compared to the standard grade selection (see
{\tt
http://asc.harvard.edu/cal/Links/Acis/acis/Cal\_prods/vfbkgrnd/}).
Real X--ray photons are practically not affected by such cleaning
(only about 2\% of them are rejected, independently of the energy
band, provided there is no pileup).

We also apply the CTI correction (see {\tt
http://cxc.harvard.edu/ciao/threads/acisapplycti/} when the
temperature of the Focal Plane at the time of the observation was 153
K.  This procedure allow us to recover the original spectral
resolution partially lost because of charge transfer inefficiency
(CTI).  The correction applies only to ACIS--I chips, since the
ACIS-S3 did not suffer from radiation damage.

If the data are taken in the FAINT mode, we run the tool {\tt
acis\_process\_events} only to apply the CTI correction and update the
gain file with the latest version provided within CALDB to date
(version 2.18).  From this point on, the reduction is similar for both
the FAINT and the VFAINT exposures.  The data are filtered to include
only the standard event grades 0, 2, 3, 4 and 6.  We checked visually
for hot columns left from the standard cleaning.  We needed to remove
hot columns by hand only in a few cases.  We identify the flickering
pixels as the pixels with more than two events contiguous in time,
where a single time interval was set to 3.3 s.  For exposures taken in
VFAINT mode, there are practically no flickering pixels left after
filtering out the bad events.  We perform a final cleaning by
filtering time intervals with high background by performing a
3--$\sigma$ clipping of the background level using the script {\tt
analyze\_ltcrv} (see {\tt
http://cxc.harvard.edu/ciao/threads/filter\_ltcrv/}).  The removed
time intervals always amount to less than 5\% of the nominal exposure
time for ACIS--I chips.  We note however, that some observations show
large flares on the ACIS--S3 chip, which are not removed by our
analysis since they are not visible on the ACIS-I chips.  We remark
that our spectral analysis is not affected by such flares, since we
always compute the background from the same observation (see below),
thus taking into account any spectral distortion of the background
itself induced by the flares.

All the observed clusters are detected with very high
signal--to--noise.  The spectrum of each source is extracted from a
circular region around the centroid of the photon distribution.  For
each cluster we select the extraction radius according to the
following criteria.  For a given radius, we find the center of the
region which includes the maximum number of net counts from the source
in the 0.5--5 keV band.  Then, we compute the signal--to--noise,
repeating this procedure for a set of radii.  Finally we choose the
radius for which the 0.5--5 keV signal--to--noise is maximum.
Incidentally, we notice that in all the cases the extraction radius
defined by maximizing the S/N is roughly 3 times the core radius
measured with a beta model (for the spatial analysis of these clusters
see Ettori et al. 2003, in preparation).  In addition, the fraction of
the net counts included in our extraction regions is always between
0.80 and 0.90 of the total.  Therefore, our choice of the extraction
radius allows us to measure the global properties of the clusters
using the majority of the signal with the highest signal--to--noise
ratio, and it minimizes any scaling effect due to possible temperature
and metallicity gradients which depend on the mass scale.

For each cluster, we use the events included in the extraction region
to produce a spectrum (pha) file.  The background is always obtained
from regions of the same exposures (same chip where the source is
located).  This is possible since all the sources are less than 3--4
arcmin in extent.  The background file is scaled to the source file by
the ratio of the geometrical area.  We checked that variations of the
background intensity across the chip do not affect the background
subtraction, by comparing the count rate in the source and in the
background at energies larger than 8 keV, where the signal from the
sources is null.

The response matrices and the ancillary response matrices are computed
in a single position at the center of the cluster.  We checked that the
variation of the spectral response of the detector across the source
position on the chip does not affect the result of the fit.  When the
source counts are spread on more than one node, we repeated the analysis
node by node; however, we never found any change in the
best--fit values.  In general, the variation of the response matrix as a
function of the position on the chip is affecting the spectral fits
only for high signal--to--noise spectra, while, in our case, the
statistical errors are always dominating.

\subsection{XMM data}

In Table \ref{exposuresxmm} we list the two XMM--Newton observations
included in our analysis, using the European Photon Imaging Camera
(EPIC) PN and MOS detectors.  We use XMM data to boost the
signal--to--noise only for the most distant clusters in our current
sample.

RXJ0849+4452 is the more luminous of the two $z\simeq1.26$ clusters in
the Lynx field (Rosati et al. 1999, Stanford et al. 2001).  We have
obtained calibrated event files for the MOS1, MOS2 and PN cameras with
the XMM Standard Analysis System (SAS) routines (SASv5.3.3). Time
intervals in which the background is increased by soft proton flares
are excluded by rejecting all events accumulated when the count rates
in the 10--12 keV band exceeds 40 cts/100s for the PN and 20cts/100s
for each of the two MOS cameras.  The final effective exposure time
amounts to 60 ks for the PN and to 112 ks for the two MOS.  Note that
we do not use data from the less luminous $z>1$ cluster in the Lynx
field, RXJ0848+4453, since it is barely detected in the XMM data due
to the high background and confusion with faint field sources.

The cluster RXJ1053.7+5735 (Hashimoto et al. 2002), is located in the
Lockman Hole field, which was observed with XMM during the Performance
Verification phase of EPIC (Jansen et al. 2001; Hasinger et al. 2001;
Mainieri et al. 2002).  Here we use a redshift of $z=1.15$ from
Hasinger et al. (2003 in preparation) which revises the initial value
published by Thompson et al. (2001).  The EPIC cameras were operated
in the standard full--frame mode.  The thin filter was used for the PN
camera, and both thin and thick filters for the MOS1 and MOS2 cameras.
The final effective exposure time, after removing background flares,
is approximately 94 ks.  RXJ1053, which is located at an off-axis
angle of $\sim 10'$, has a double--lobed morphology with the two cores
1 arcmin apart (see Hashimoto et al. 2002).  We will present the
analysis of the whole cluster, as well as the separate analyses of the
two clumps, which show significantly different temperatures.

\section{RESULTS}

\subsection{Spectral analysis}

The spectra are analyzed with XSPEC v11.2.0 (Arnaud 1996) and fitted
with a single temperature MEKAL model (Kaastra 1992; Liedahl et
al. 1995), where the ratio between the elements are fixed to the solar
value as in Anders \& Grevesse (1989).  These values for the solar
metallicity have recently been superseded by the new values of
Grevesse \& Sauval (1998), who use a 0.676 times lower Fe solar
abundance.  However, we prefer to report metallicities in units of the
Anders \& Grevesse abundances since most of the literature still
refers to these old values.  Since our metallicity depends only on the
Fe abundance, updated metallicities can be obtained simply by
rescaling by 1/0.676 the values reported in Table 3.  We model the
Galactic absorption with {\tt tbabs} (see Wilms, Allen \& McCray
2000).

Before performing the fits, we apply a double correction to the {\sl
arf} files.  We apply the script {\tt apply\_acisabs} by Chartas and
Getman to take into account the degradation in the ACIS QE due to
material accumulated on the ACIS optical blocking filter since launch
(see {\tt http://cxc.harvard.edu/ciao/threads/apply\_acisabs/}).  This
correction applies to all the observations.  For data taken with
ACIS--I, we also manually apply a correction to the effective area
consisting in a 7\% decrement below 1.8 keV to homogenize the
low--energy calibrations of ACIS--S3 and ACIS--I (see Markevitch \&
Vikhlinin 2001).

The fits are performed over the energy range 0.6--8 keV.  We exclude
photons with energy below 0.6 keV in order to avoid systematic biases
in temperature determination due to uncertainties in the ACIS
calibration at low energies.  The effective cut at high energies can
be lower than 8 keV, since the signal--to--noise for thermal spectra
rapidly decreases at energies above 5 keV.  Generally, we cut at 7--8
keV after visually inspecting each spectrum.  The excluded energy
range (0.3--0.6 keV for Chandra, and below 0.5 keV for XMM) does not
affect the determination of the metallicity, which is dominated by the
Fe K line complex at rest--frame energies of 6.4--6.7 keV (always
located above $\sim$3 keV for our cluster sample).  All the iron and
oxygen L spectral features at 1 keV rest--frame are redshifted below
0.8 keV for $z\magcir 0.4$, i.e. for the majority of our sample.  We
note that our Fe K-line diagnostic is simpler and more robust than
that based on the line--rich region around 1 keV, where the line
emission is dominated by the L--shell transition of Fe, and the
K--shell transitions of O, Mg, and Si.

We use three free parameters in our spectral fits: temperature,
metallicity and normalization.  We freeze the local absorption to the
galactic neutral hydrogen column density ($N_H$, table \ref{results}), as
obtained from radio data (Dickey \& Lockman 1990), and the redshift to
the value measured from the optical spectroscopy.  Since both the
absorption and the redshift can be in principle determined from the
X--ray data itself, we checked that our results do not change when
these parameters are left free, as described below.

A well-known strong degeneracy exists between $N_H$ and the best--fit
temperature: a higher $N_H$ implies always a lower $kT$, with a rough
scaling $N_H\propto T^{-4}$.  We looked for a possible biases by
comparing the best--fit values obtained after fixing the local
absorption to the Galactic one, with those obtained using $N_H$ as a
free parameter.  An inspection of the contour plots in the Figures of
Appendix A, shows that the Galactic values by Dickey \& Lockman (1990)
are typically recovered within the $1 \sigma$ contours in the
$N_H$--$T$ space, with three exceptions (MS2137, MS0451, MS1054), for
which a discrepancy larger than $3 \sigma$ is found.  In addition, we
also notice that the best--fit temperature is found in 3 cases
somewhat higher than the value obtained for a free $N_H$, in 4 cases
somewhat lower, while in the remaining 11 cases it is practically
coincident.  Therefore, we conclude that by fixing the local
absorption to the Dickey \& Lockman values, we do not introduce any
significant systematic bias in the measure of the temperature.

We also checked that the best--fit temperature remains unaffected when
the redshift is left free.  The best--fit redshift is consistent with
the spectroscopic redshift for most of the clusters within $1 \sigma$,
and between $1$ and $2 \sigma$ in the remaining 3 cases (RXJ1113,
RXJ1350, RXJ0849).  Generally, small changes in the redshift can give
significant changes in the best--fit metallicity, since its value
depends critically on the position of the Fe line.  We find that the
metallicity tends to be higher when the redshift is left free. This is
due to positive noise fluctuations which systematically tend to boost
the Fe line equivalent width.  It is evident that for such low
signal--to--noise spectra the best strategy is to lock the cluster
redshift to the spectroscopic value.

In performing the spectral fits we used the Cash statistics applied to
source plus background (see {\tt
http://heasarc.gsfc.nasa.gov/docs/xanadu/xspec/manual/node57.html}),
which is preferable for low signal--to--noise spectra (Nousek \& Shue
1989).  We also performed the same fits with the $\chi^2$ statistics
(with a standard binning with a minimum of 20 photons per energy
channel in the source plus background spectrum) and verified that all
the best--fit models have a reduced $\chi^2 \sim 1$.

\subsection{Single source analysis}

In this section, we present the results of the spectral analysis of
the 18 clusters.  We remind that RXJ0152 shows two distinct cores
which we treat as separate objects.  As for RXJ1053, the spatial
separation of the two clumps is less clear.  Therefore, we show the
analyses of the whole cluster as well as that of the two clumps,
which, indeed, have significantly different temperatures (Hashimoto et
al. 2002).  The results are listed in Table \ref{results}.

\begin{figure}
\centerline{\psfig{figure=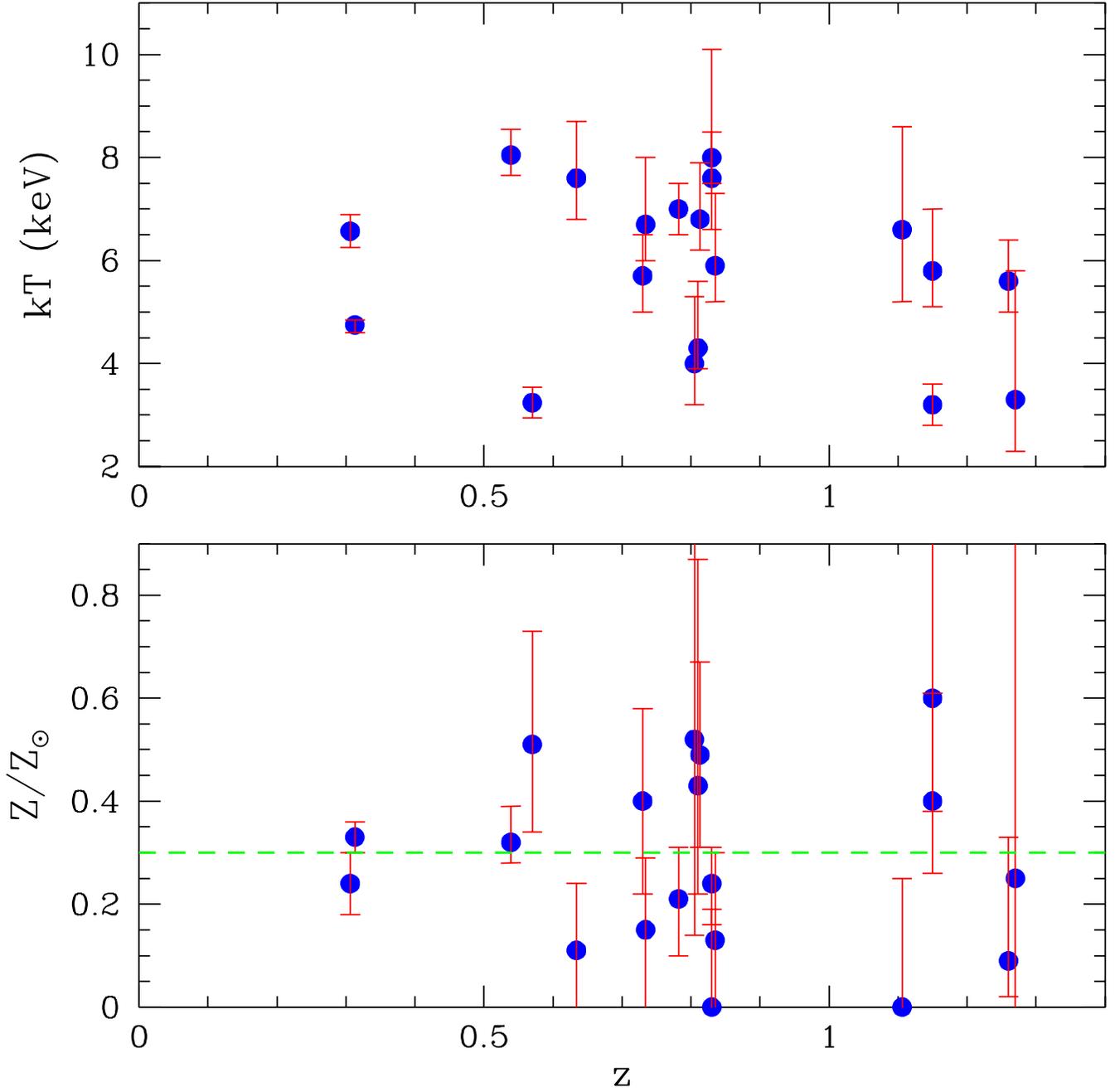,height=8.in,width=8.in}}
\caption{Temperature (upper panel) and Fe abundance (lower panel) vs
redshift for the 18 clusters in the sample.  Error bars refer to 1
$\sigma$ c.l. computed for one interesting parameters. The metallicity
is given in units of the solar Fe abundance as measured by Anders \&
Grevesse (1989)
\label{tm_vs_z}}
\end{figure}

\begin{figure}
\centerline{\psfig{figure=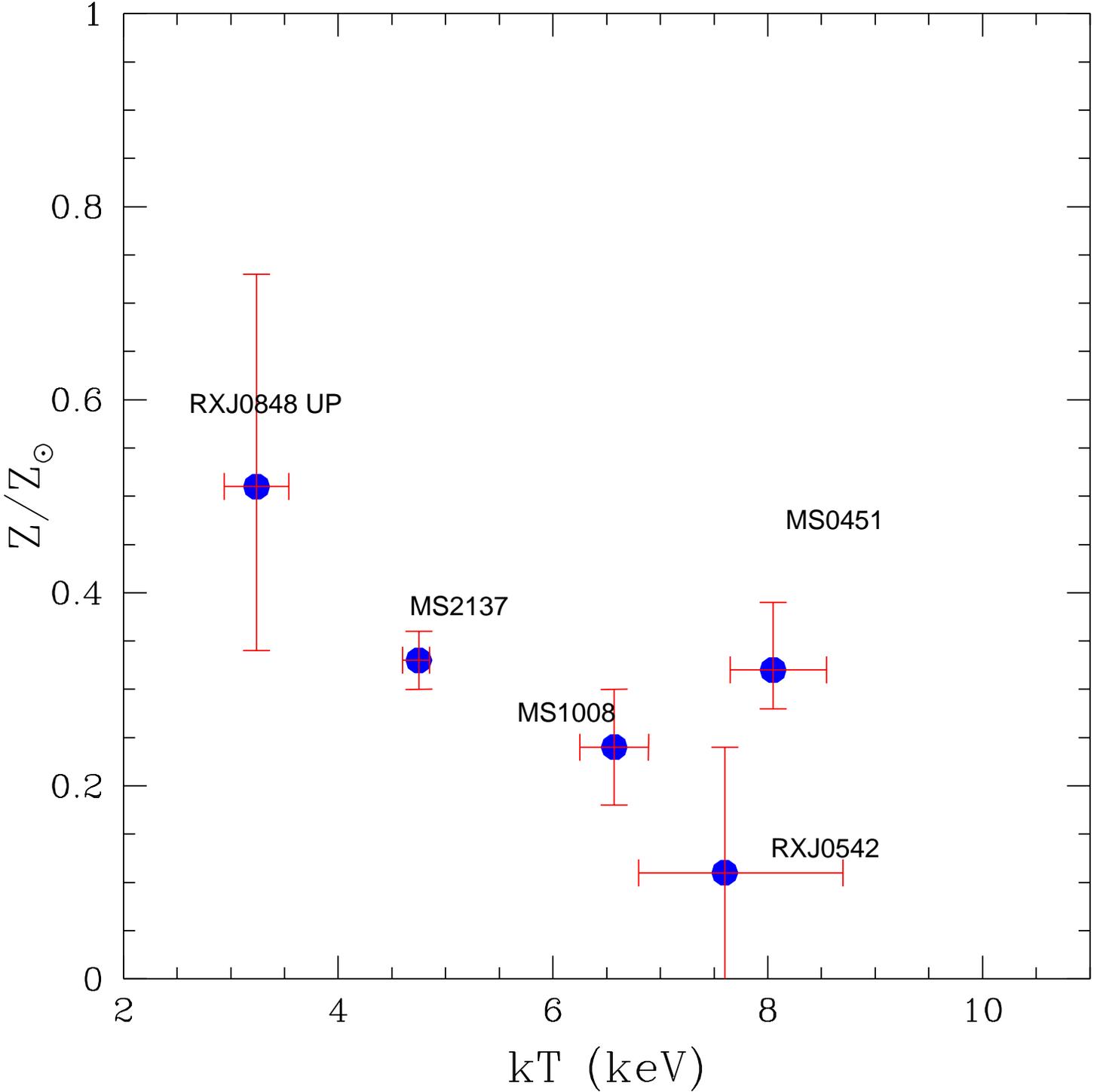,height=8.in,width=8.in}}
\caption{Temperature--metallicity plot for 5 clusters in the redshift
range $0.3 < z < 0.65$.  Error bars refer to 1 $\sigma$ c.l. for one
interesting parameter.  The metallicity is given in units of the solar
Fe abundance as measured by Anders \& Grevesse (1989).
\label{met1}}
\end{figure}

\begin{figure}
\centerline{\psfig{figure=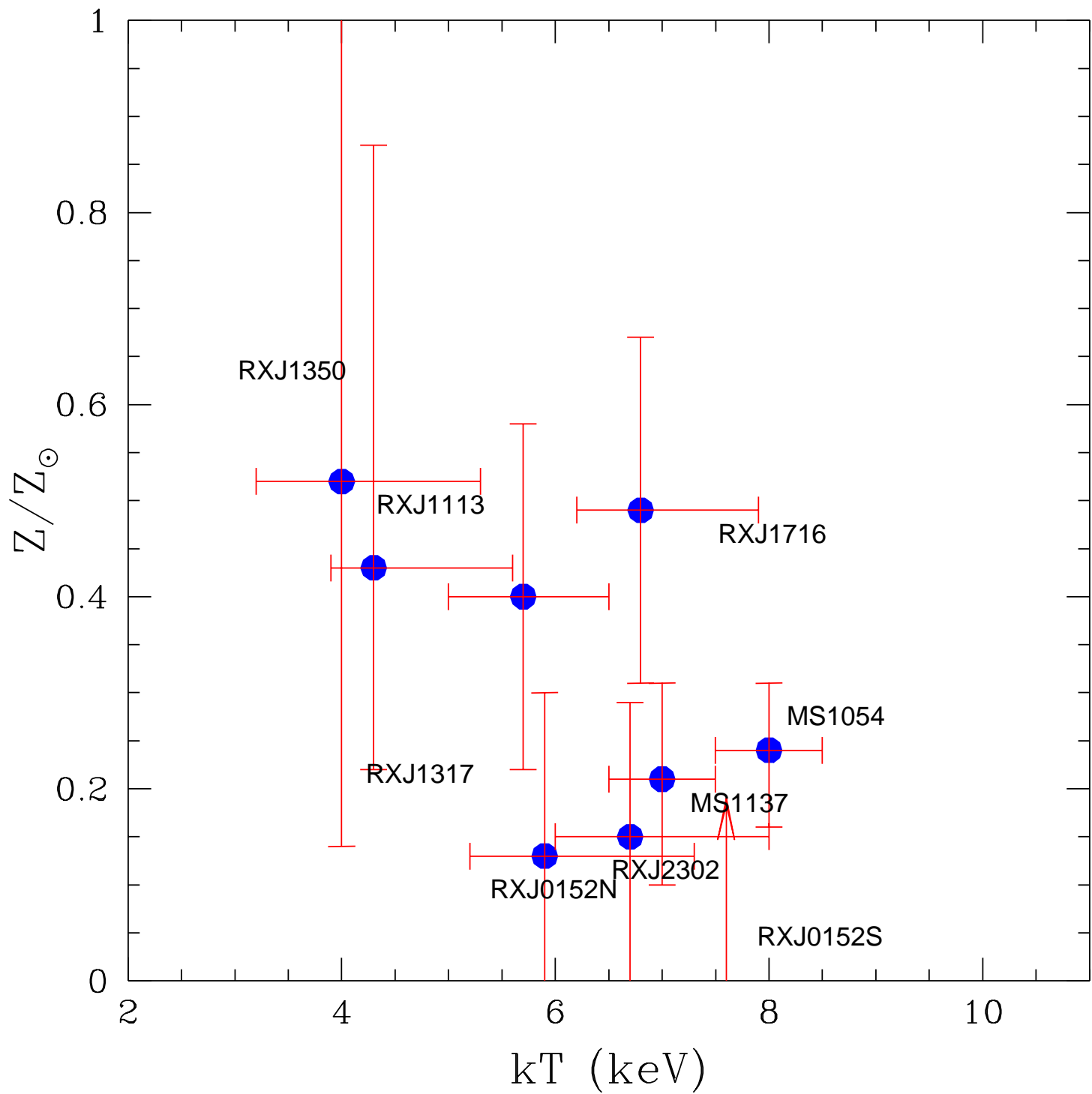,height=8.in,width=8.in}}
\caption{Same as in Figure 2 for 9 clusters in the redshift
range $0.65<z<0.90$.  
\label{met2}}
\end{figure}

\begin{figure}
\centerline{\psfig{figure=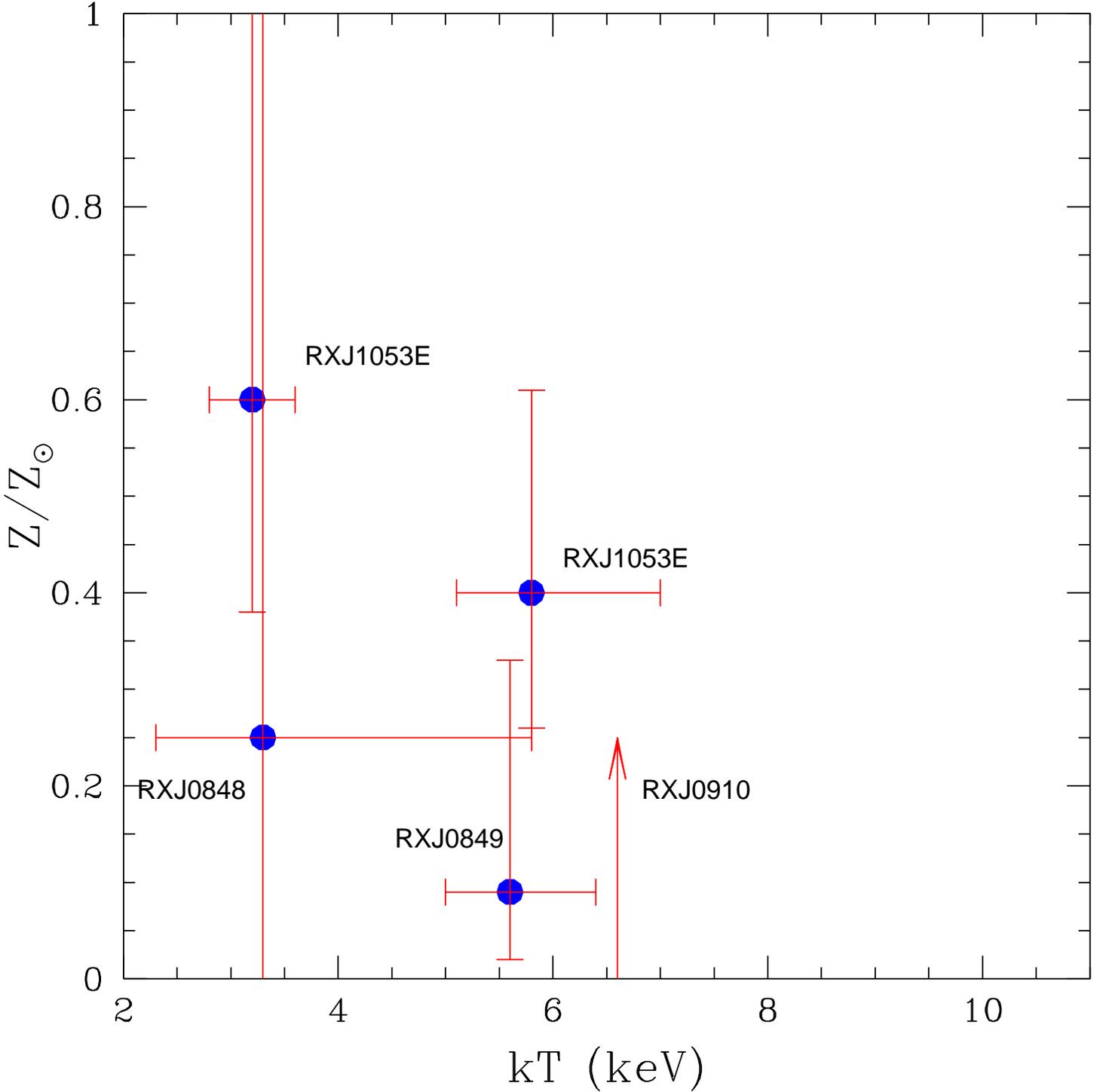,height=8.in,width=8.in}}
\caption{Temperature--metallicity plot in the redshift range $1.1 < z
< 1.3$ for the 4 clusters observed with Chandra and XMM.  We obtain a
positive detection of the Fe abundance only for RXJ1053 (observed only
with XMM), and upper limits for the other three clusters.  Error bars
refer to 1$\sigma$ c.l. for one interesting parameter.  The
metallicity is given in units of the solar Fe abundance as measured by
Anders \& Grevesse (1989).
\label{met3}}
\end{figure}

In Figure \ref{tm_vs_z} we show the distribution of temperature and Fe
abundance for our sample as a function of redshift (error bars are at
$1 \sigma$ confidence level).  We looked for possible correlations
between different parameters.  For the temperature--redshift relation
we find a Spearman's rank coefficient of $r_s =-0.20$ for 19 degrees
of freedom, thus consistent with no correlation.  As for the
metallicity--redshift relation, we find again no correlation, with
Spearman's rank coefficient of $r_s =-0.15$ for 19 degrees of freedom,

We show our results for three redshift intervals: $0.3<z<0.65$ (five
objects, Figure \ref{met1}), $0.65<z<0.90$ (nine objects, Figure
\ref{met2}), $0.90<z<1.25$ (five objects, Figure \ref{met3}).  We
notice a large intrinsic scatter, which is, however, comparable with
the statistical errors.  Given the small size and the limited
temperature range of our sample, it is difficult to draw any
conclusion on the relation between Fe abundance and temperature at a
given redshift.  We notice, however, a trend of lower Fe abundance at
higher temperatures ($kT > 5$ keV) with a Spearman's correlation
coefficient $r_s=-0.57$ for 19 degree of freedom, slightly below the
90\% confidence level, for the whole sample.  We find a Spearman's
correlation coefficient of $r_s = -0.70$, $r_s = -0.50$, and
$r_s=-0.70$ for 5, 9 and 5 degrees of freedom respectively in the three
redshift intervals.  We note that the presence of such a correlation
is in keeping with observations of local and moderate redshift
clusters (see Mushotzky \& Loewenstein 1997; Finoguenov, Arnaud \&
David 2001), which suggest higher Fe abundances for lower temperature
systems.  Given this result, we choose to differentiate our sample in
low and high temperature clusters.  If we compute the Fe
abundance--redshift relation for the 13 clusters with $kT>5$ keV, we
find a Spearman's rank coefficient of $ r_s=-0.40$ for 13 degrees of
freedom, which at face value implies a correlation at $\mincir 2
\sigma$ confidence level.

In the intermediate redshift bin ($0.65<z<0.90$), the Fe emission line
is clearly detected in the average cluster population (see also \S
3.3), although the metallicity is consistent with zero (within 1
$\sigma$) for three clusters, and for one of them (RXJ0152S) we have
only an upper limit.  In the highest redshift bin ($z>1$, see Figure
\ref{met3}), the Fe K line is detected only in the spectrum of RXJ1053
(for which $Z>0.1 \, Z_\odot$ at the 90\% c.l.) as well as in the two
separate clumps, which have different temperatures: $kT = 3.2\pm 0.4$
keV for the eastern clump and $kT = 5.8^{+1.2}_{-0.7}$ keV for the
western clump.  In the spectra of the other two clusters with $kT> 5$
keV we do not detect the iron line.  To investigate further the Fe
abundance--redshift relation, we propose to measure the {\sl average}
metallicity of the ICM as a function of cosmic epoch by combining
photons from subsamples of clusters as described below.

\subsection{Combined spectral analysis}

A robust measurement of the {\sl average} ICM metallicity as a
function of cosmic epoch can be obtained by combining the information
contained in the individual spectra, after grouping them in a number
of redshift bins.  This technique is similar to the stacking analysis
often performed in optical spectroscopy, where spectra from a
homogeneous class of sources are averaged together to boost the
signal--to--noise, thus allowing the study of otherwise undetected
features. In this case, one cannot stack different spectra due to
their different shapes (different temperatures), however one can
perform a {\sl combined fit}, as explained below.

\begin{figure}
\centerline{\psfig{figure=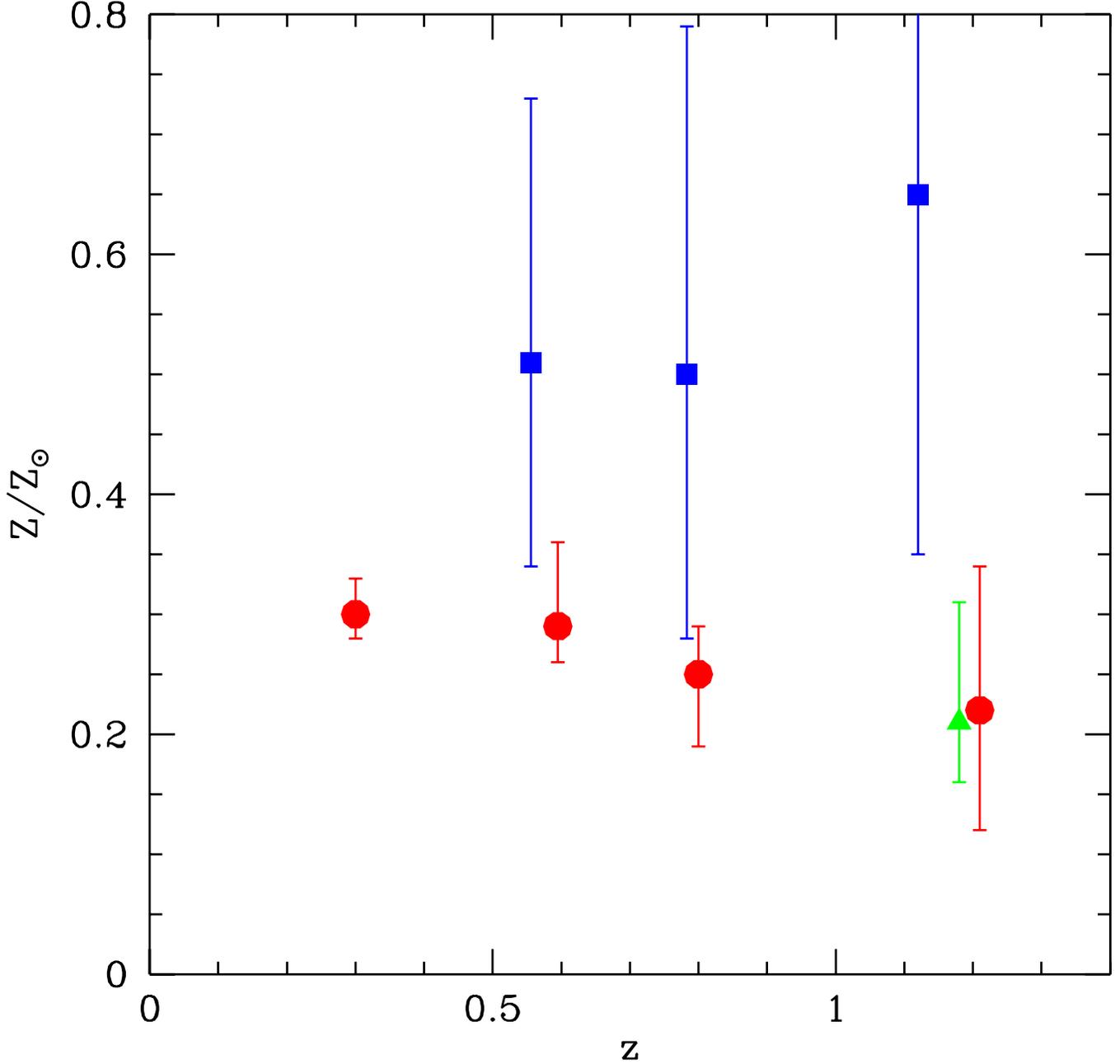,height=8.in,width=8.in}}
\caption{Average metallicity as a function of the redshift for the
four bins defined in the text, for clusters with $kT>5$ keV (solid
circles) and $kT<5$ keV (solid squares).  For the highest redshift bin
($z\sim 1.2$), we have only an upper bound from the three clusters
with $kT>5$ keV, while we measure the iron abundance only for RXJ1053
($kT\simeq 4 $ keV).  The triangle is the combined fit of all the
clusters at $z>1$ irrespective of the temperature.  Error bars refer
to $1 \sigma$ c.l.
\label{avmet2}}
\end{figure}

We define 4 redshift bins as follows.  The first bin includes only
MS1008 and MS2137 with $\langle z \rangle =0.31$; the second one
includes MS0451, RXJ0848$-$4456 and RXJ0542, with $\langle z \rangle =
0.57$; the third one includes 9 clusters: RXJ2302, RXJ1113, MS1137,
RXJ1317, RXJ1350, RXJ1716, MS1054, RXJ0152N, RXJ0152S, with $\langle z
\rangle = 0.80$; the fourth one includes the five objects at $z>1$:
RXJ0910, RXJ0849 (observed with Chandra and XMM), RXJ0848, and the two
clumps of RXJ1053 in the Lockman Hole observed only with XMM.  The
average redshift in the highest--z bin is $\langle z \rangle \simeq
1.2$.  In each redshift bin, we performed a combined fit, leaving
temperature and normalization free for each object, using however a
single metallicity parameter for all the clusters in the subsample.
We also subdivide the clusters within each redshift bin according to
temperature, below and above 5 keV, in order to investigate whether
either low-- or high--temperature clusters contribute to a possible
evolution of the average metallicity.  We show our results in Figure
\ref{avmet2} (the best fit values are listed in Table \ref{tableZ}).

We measure a mean metallicity very close to the canonical local value
of $0.3 Z_\odot$ out to $z\simeq 0.6$ for high temperature clusters.
We have only one cluster with temperature below 5 keV at $z<0.6$
(RXJ0848+4456), for which we find $Z\sim 0.5 \, Z_\odot$ with large
error bars.  More interestingly, we obtain for the first time a robust
determination of the Fe mean abundance at $z=0.8$: $\langle Z\rangle =
0.25^{+0.04}_{-0.06}$ in clusters with $kT> 5$ keV, and possibly
higher ($Z\simeq 0.5 \, Z_\odot$) for lower temperature clusters.  At
$z>1$ the combined fit for the three $kT>5$ keV clusters yields a weak
positive detection of $Z = 0.22\pm 0.12\, Z_\odot$.  This measure is
consistent with the local value within one sigma.  The two measures
for with $kT<5$ keV (RXJ0910 and the eastern clump of RXJ1053) give a
very large value for the iron abundance of $Z\simeq 0.6 \, Z_\odot$,
but with large uncertainties.  If we combine all the clusters at $z>1$
irrespective of the temperature, we obtain $\langle Z\rangle =
0.21^{+10}_{-0.05}$.  This measure is dominated by RXJ1053, which
contributes to the combined fit with 870 total net counts, while we
have 1550 net counts from the remaining three clusters.  This measure
of the average iron abundance can be considered an average among
clusters with different properties: if the other high--z clusters had
the same metallicity, we would have been able to measure it.

The values of mean metallicity as a function of redshift can be fitted
with a constant value: a $\chi^2$ minimization on the four points
obtained for clusters with $kT>5$ keV gives $Z=0.286 Z_\odot$ as the
best choice ($\chi^2 \sim 3$ for 3 degrees of freedom).  As a result,
we do not find significant evolution in the mean ICM metallicity out
to $z\sim\! 1.2$.  The apparent decline of $\langle Z\rangle$ at $z>1$
in high--temperature clusters (Figure \ref{avmet2}) has a very low
confidence level (between 70\% and 80\%).

\begin{figure}
\centerline{\psfig{figure=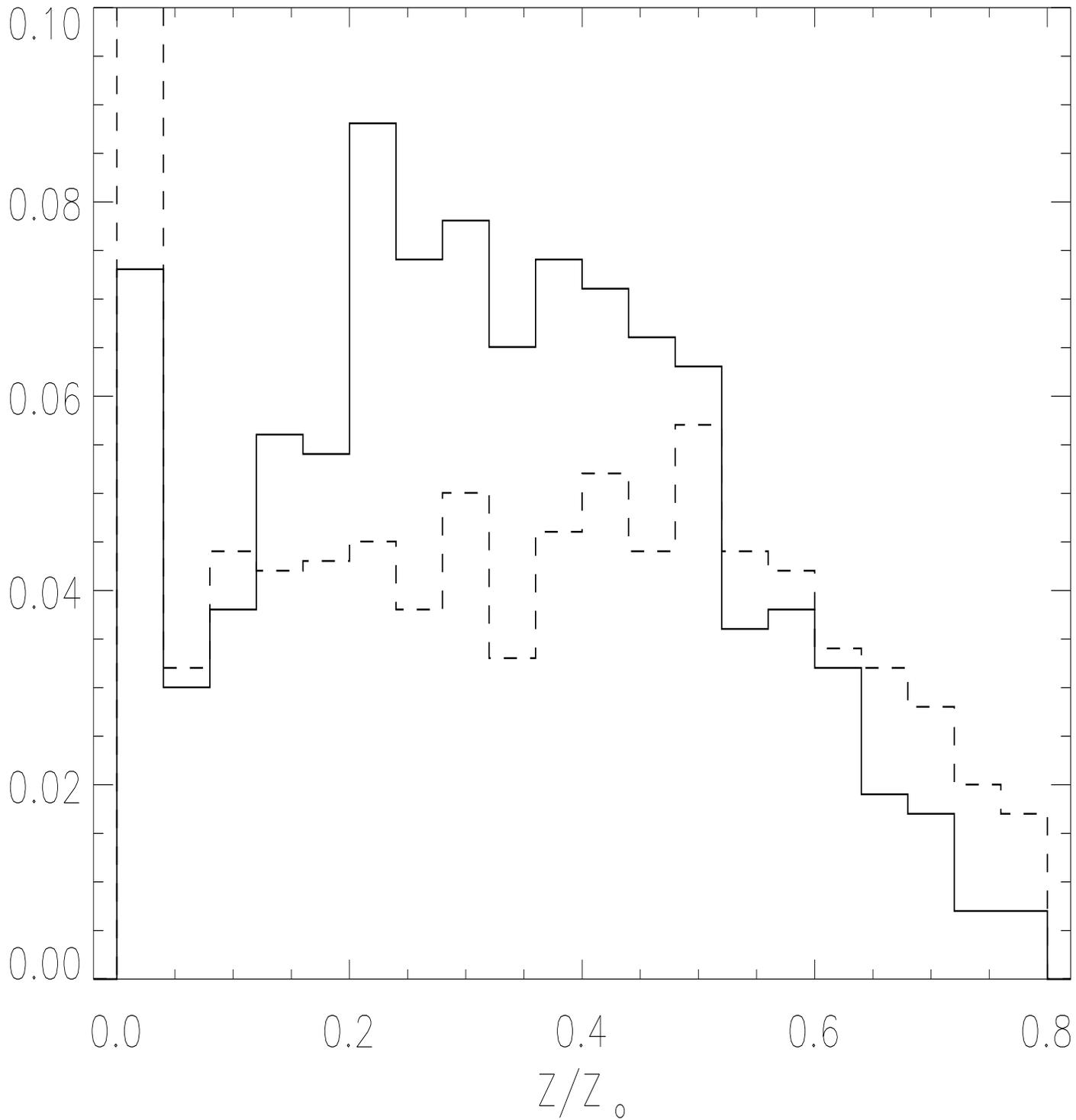,height=8.in,width=8.in}}
\caption{Distribution of the best--fit metallicity values from the
combined fit of the 4 clusters in our sample at $z>1$, when
simulated $10^3$ times (solid line).  The input metallicity is $0.3
Z_\odot$. The dashed line is the result of the simulations when only
the three clusters observed with Chandra are considered.
\label{met_hist}}
\end{figure}

To better establish the statistical significance of our Fe abundance
measurement in the highest redshift bin, as well as possible biases on
its best--fit value, we performed simulations of the combined spectral
analysis by creating synthetic spectra with the same signal--to--noise
and global parameters ($L_X, T, N_H, z$) as our high--z clusters.  We
fixed the metallicity to be the same, $ 0.3 Z_\odot$, in the 5 objects
at $z>1$.  We used XSPEC ({\tt fakeit} command) to simulate source
spectra and background spectra for each cluster starting from the
actual value of the background derived from the data.  We then ran
1000 simulation for each cluster and analyzed them with the same
fitting procedure adopted on the real data to derive the average
metallicity at redshift $z=1.2$ (we recall that the combined fit is
performed on 6 separate spectra since for RXJ0849 we use 3 separate
data sets: Chandra-ACIS, XMM--PN and XMM--MOS). We verified that the
best--fit values for the temperatures for each cluster are distributed
around their respective input values.  In Figure \ref{met_hist}, we
show the distribution of the best--fit values for the metallicity from
the combined fit of 1000 simulations (solid line).  The distribution
is rather broad, with a spike at zero metallicity (in 5\% of the
case), underlying the difficulty in constraining the metallicity at
such high redshifts from an effective number of $\sim\! 2400$ photons.
The median value of the recovered best--fit metallicity is $Z=0.325\,
Z_\odot$; the 84 lower percentile corresponds to $Z=0.13\, Z_\odot$,
thus including our measured value of $Z\simeq 0.20\, Z_\odot$.  In the
simulations reproducing only the observations of clusters with $kT>5$
keV (dashed line) the 84 lower percentile corresponds to $Z=0.02\,
Z_\odot$.  These simulations show that in order to detect possible
evolution of the average metallicity at $z>1.2$ in high temperature
clusters, we need to enlarge substantially our sample.

\section{DISCUSSION AND CONCLUSIONS}

We presented the spectral analysis of 18 clusters of galaxies at
intermediate--to--high redshifts observed by Chandra and XMM.  Our
analysis represents the first attempt to date to follow the evolution
of iron content of the ICM out to $z\magcir 1$.

As a first firm result of our analysis, we determine the average ICM
metallicity with a 20\% uncertainty at $z<1$ ($Z/Z_\odot =
0.25^{+0.04}_{-0.06}$ at $\langle z\rangle = 0.8$) and find no
evidence of evolution over this redshift range.  This finding is
consistent with the expectation that the peak of star formation in
proto--cluster regions occur at redshift $z\simeq 3$--4. The extension
of this analysis at $z>1$, with similar accuracy in the determination
of the Fe abundance, will be very valuable and could further constrain
the epoch at which the ICM was metal--enriched from the last
significant episode of star formation.

Our sample includes only 4 clusters at $z>1$ to date.  One of these
clusters, RXJ1053, shows two clumps with different temperatures. In
this way, we end up with three structures having $kT>5$ keV and two at
lower temperature.  A combined fit across the high temperature systems
yields $Z = 0.22\pm 0.12$ for the average metallicity, still
consistent with zero metallicity at the 90\% c.l.. Only after
combining all our clusters at $z>1$, irrespective of their
temperature, we obtain a more robust detection of metallicity, $Z =
0.21^{+0.12}_{-0.05} Z\odot$. This result suggests no evolution of the
mean iron abundance out to $z\simeq 1.2$.

Our analysis demonstrates that measurements of the ICM metallicity at $z>1$
are indeed possible with the current generation of X-ray satellites
and with reasonable exposure times. A more precise statement of a
possible evolution pattern requires that we can at least double the
number of secured clusters at $z>1$. Only one additional RDCS cluster
at $z=1.24$ remains to be observed with Chandra and XMM.  However,
distant clusters are expected to be discovered with serendipitous
on--going surveys, which take advantage of Chandra and XMM archival
pointings (e.g., Boschin 2002). Precise measurements of the metal
content of clusters at large look-back times are crucial for the study
of the thermodynamics of the ICM and of the star formation processes
in cluster galaxies.

Much emphasis has been given recently to the evolution of global
scaling relations for the ICM, such as the $L_X-T$ relation
(e.g. Holden et al. 2002, Vikhlinin et al. 2002). However, a knowledge
of the history of the ICM metal enrichment is also needed for
understanding the mode and epoch of cluster formation in its hot and
cold phase. In this respect, measuring the properties of the ICM at
redshifts 1--2 is mandatory to constrain the physical processes
involved in the diffusion of energy and metals within clusters.  For
example, if the evolution of metallicity is accompanied by an
evolution of other thermodynamical properties of the ICM, such as the
entropy, one could in principle establish a link between the star
formation processes, which are responsible for the enrichment, and the
non--gravitational energy injection into the ICM. On the other hand,
if the X--ray properties of the ICM at $z>1$ are not significantly
different from the local ones, then any non--gravitational heating of
the ICM and, therefore, its metal enrichment, needs to be completed by
$z=1$.

A lack of evolution of the Fe abundance at $z>1$ can be translated
into a constraint on the epoch of the star formation episode, which is
responsible for the metal enrichment.  The Fe abundance, at least in
central cluster regions, is expected to be dominated by SNIa (but see
Gastaldello \& Molendi 2002). If the bulk of the iron mass is already
in place at $z\simeq 1$, one would expect that most of the SNIa
already exploded and injected their metals into the ICM by $z\simeq
1.2$.  Assuming a lifetime of roughly 1 Gyr, this would imply that the
last significant episode of star formation occurred at redshift $z>
1.5$, thus consistent with the expectation that the peak of
star-formation in proto-cluster regions took place at $z\sim 3$.

Finally, measuring the ICM metallicity beyond $z=1$ would have
important implications on the SN rate in clusters.  For the purposes
of the discussion here, let us assume that energy input and the
metallicity input are from Type Ia SN. A canonical value for the yield
of Fe from Type Ia SN is 0.5 $M_{\odot}$. Therefore, bringing the ICM
metallicity to $\simeq 0.3\, Z_\odot$ requires $\sim 9 \times
10^{9}M_\odot$ of Fe from a cluster of mass $\sim 10^{14}M_{\odot}$
and thus the total number of Type Ia SN is approximately $2 \times
10^{10}$. If these Type Ia SN were produced between redshift 1 and 2,
then their rate would be of order $\sim 10$ yr$^{-1}$.  The physical
mechanism for Type Ia SN is still unclear but a canonical delay of 1-3
Gyr puts the formation epoch of the parent stellar population at
$z\geq 3$. Therefore, monitoring the SN rate in these high redshift
clusters could produce important physical constraints.

\acknowledgements

The authors thank Pasquale Mazzotta and Alexei Vikhlinin for useful
insights in data reduction and analysis, Yasuhiro Hashimoto for
providing the XMM--MOS spectra of RXJ1053, as well as Sabrina De
Grandi for helpful discussions.  We also thank the entire Chandra Team
for continuous support in handling the data.  P. Tozzi acknowledges
support under the ESO visitor program in Garching during the
completion of this work.

\newpage

\section*{Appendix A: Single Sources analysis}

\begin{figure}
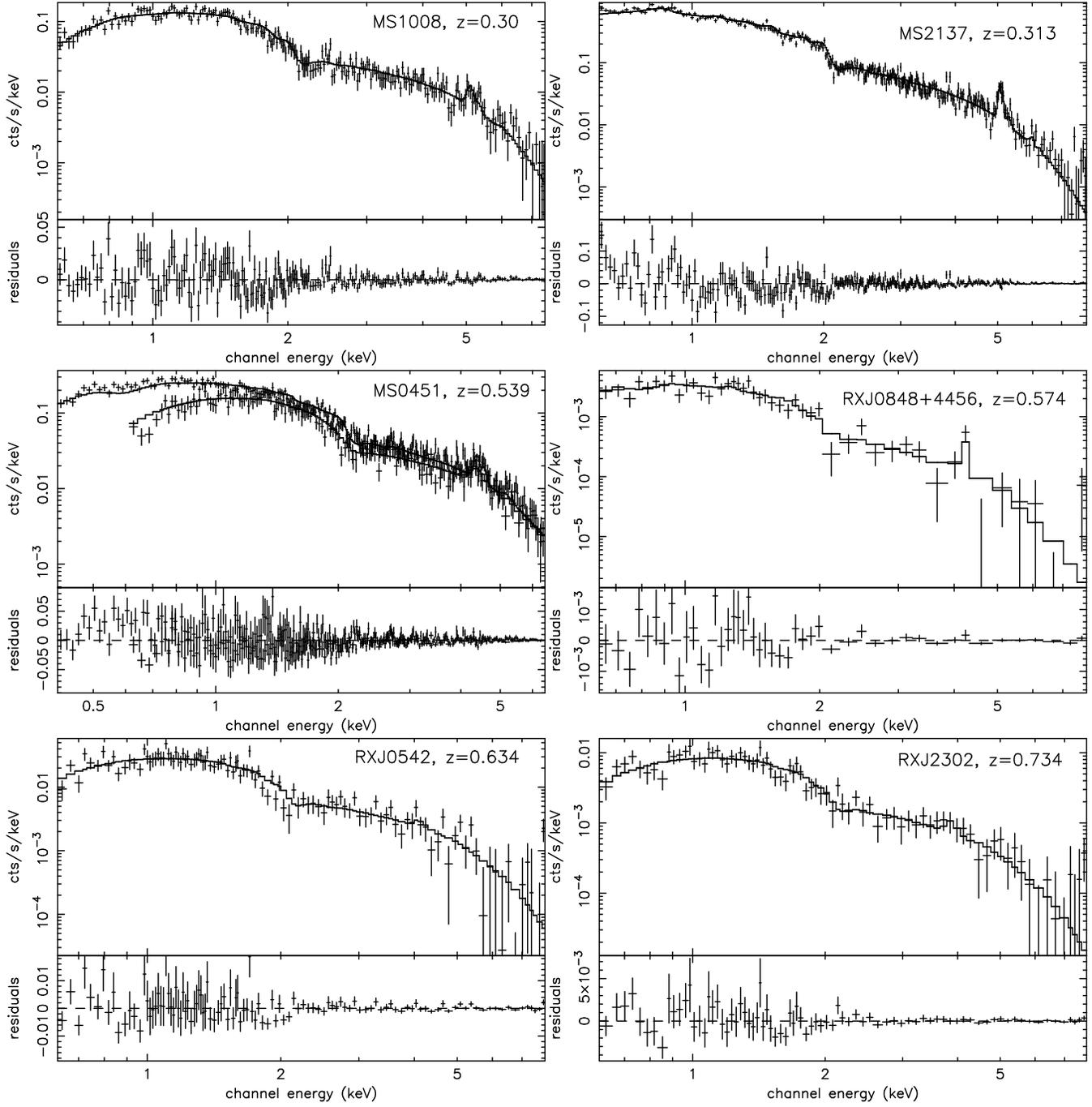
 

\centerline{\psfig{figure=cl_926_data.ps,angle=270,height=6cm}\psfig{figure=cl_928_data.ps,angle=270,height=6cm}}

\centerline{\psfig{figure=cl_529_data.ps,angle=270,height=6cm}\psfig{figure=cl2_up_data.ps,angle=270,height=6cm}}

\centerline{\psfig{figure=cl_914_data.ps,angle=270,height=6cm}\psfig{figure=cl_918_data.ps,angle=270,height=6cm}}

\caption{The spectra of the clusters observed with Chandra with the
folded thermal best--fit model (continuous lines).  The lower panels
show the residuals.  Clusters are sorted by increasing redshift.  Note
that in the case of MS0451 we show the data for ACIS--I and for ACIS--S.}
\label{spectra1} 
\end{figure}

\begin{figure} 

\centerline{\psfig{figure=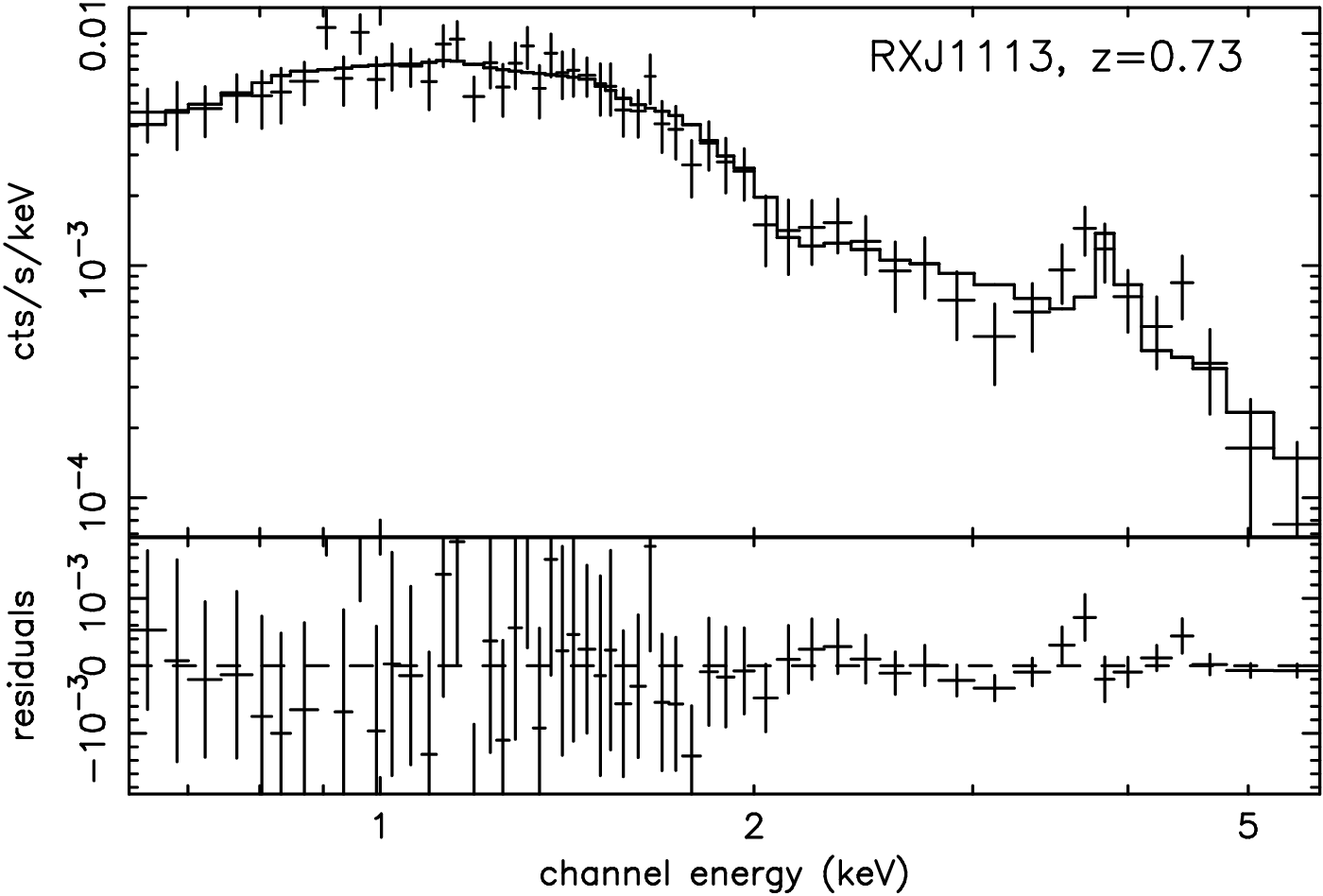,angle=270,height=6cm}\psfig{figure=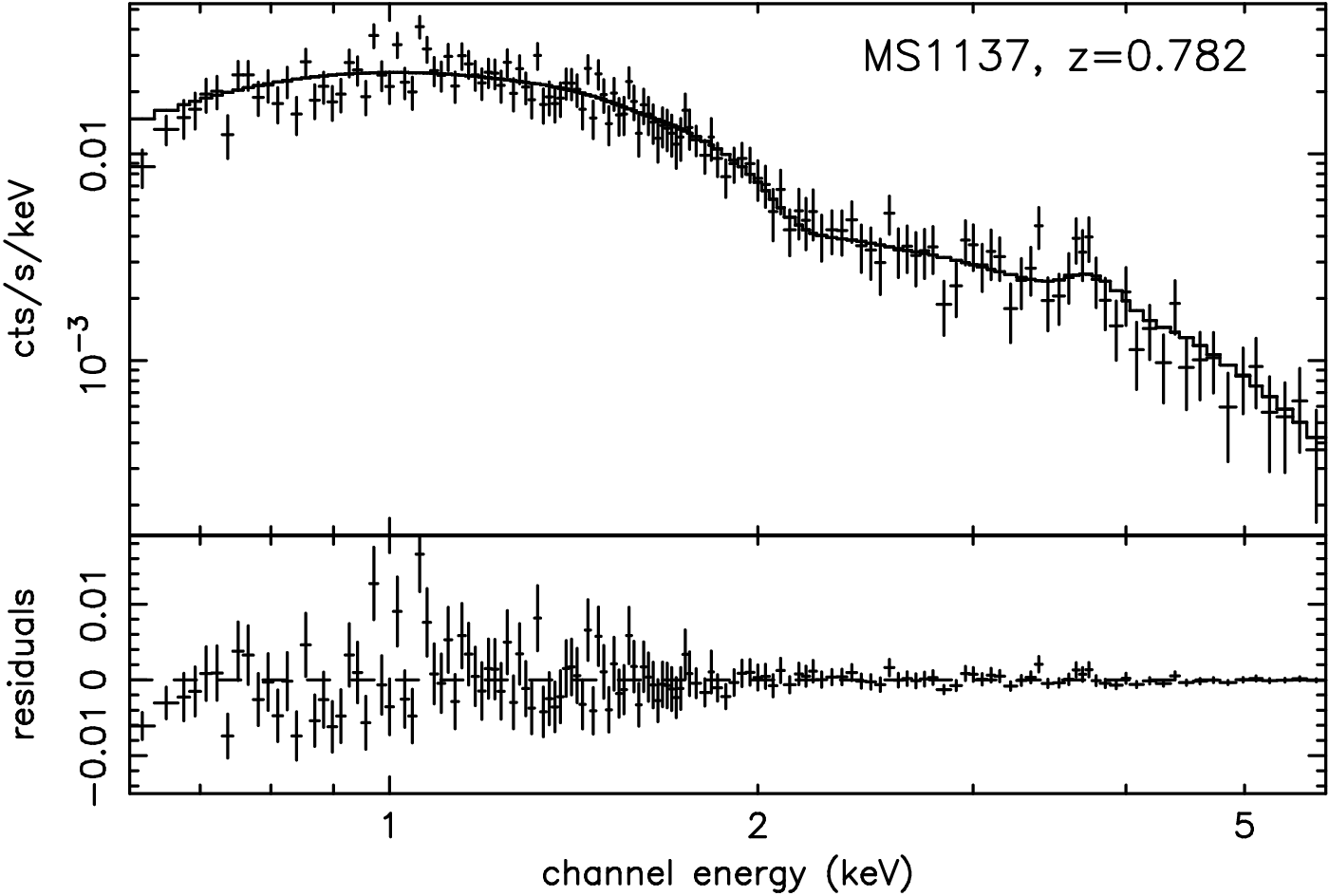,angle=270,height=6cm}}

\centerline{\psfig{figure=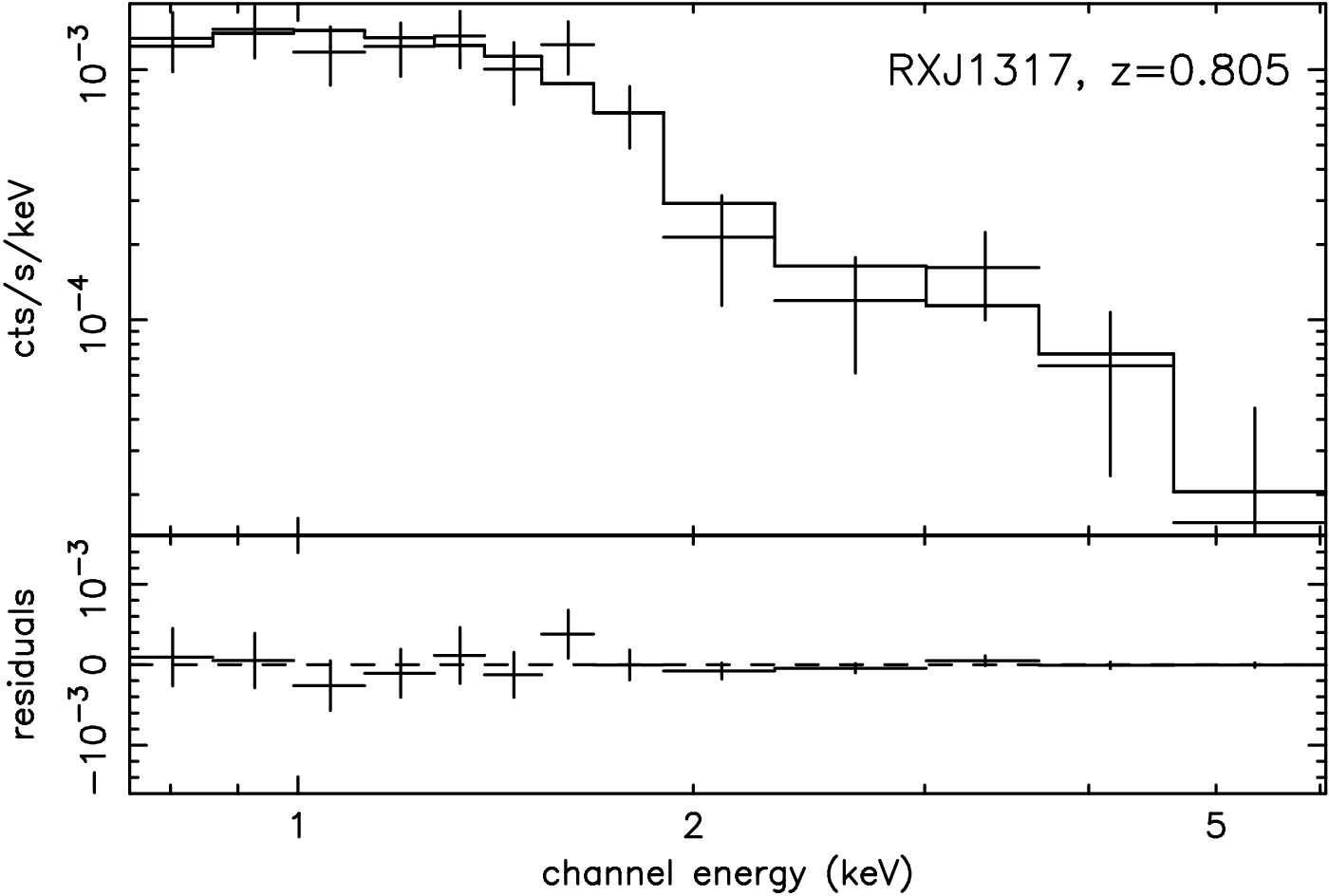,angle=270,height=6cm}\psfig{figure=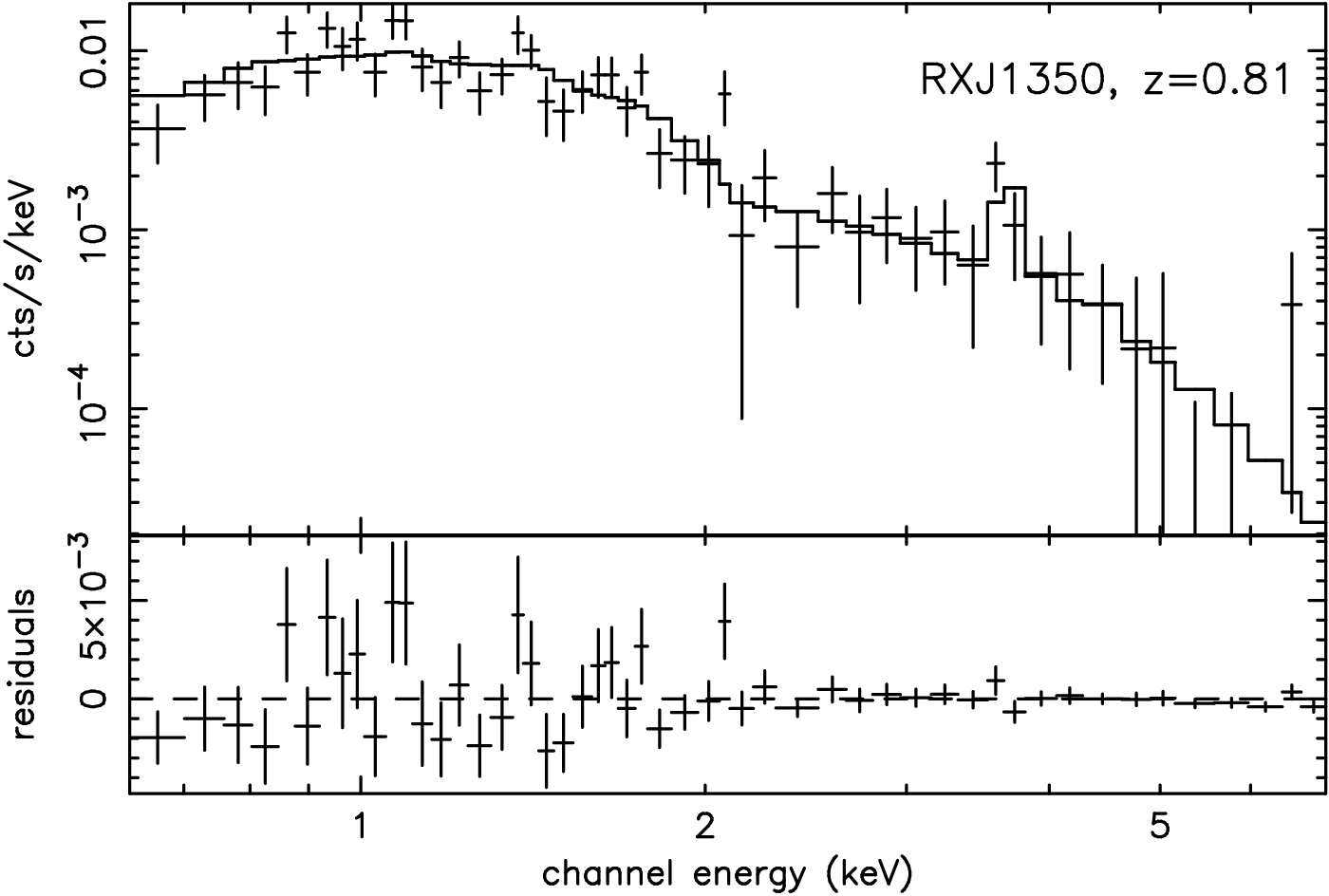,angle=270,height=6cm}}

\centerline{\psfig{figure=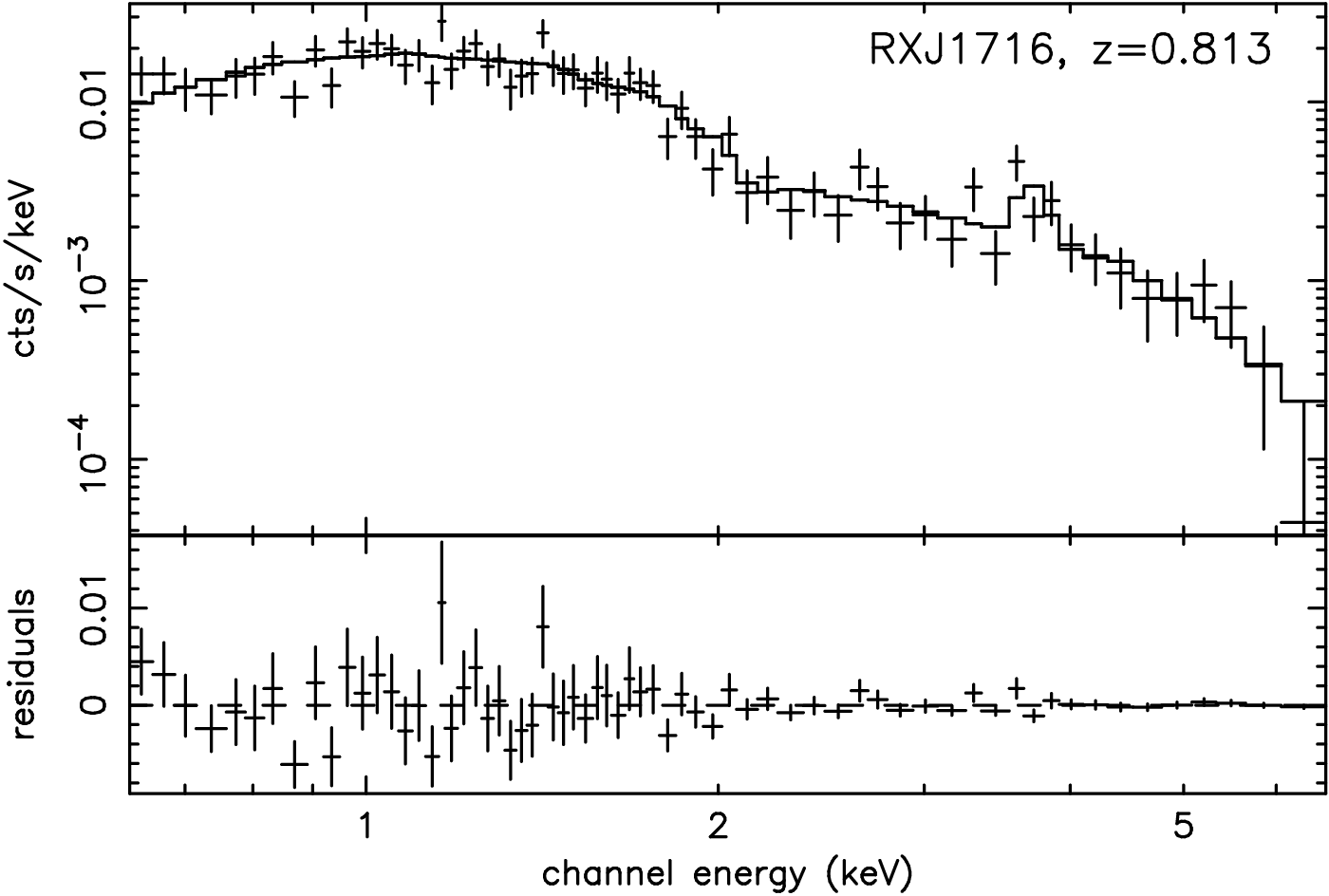,angle=270,height=6cm}\psfig{figure=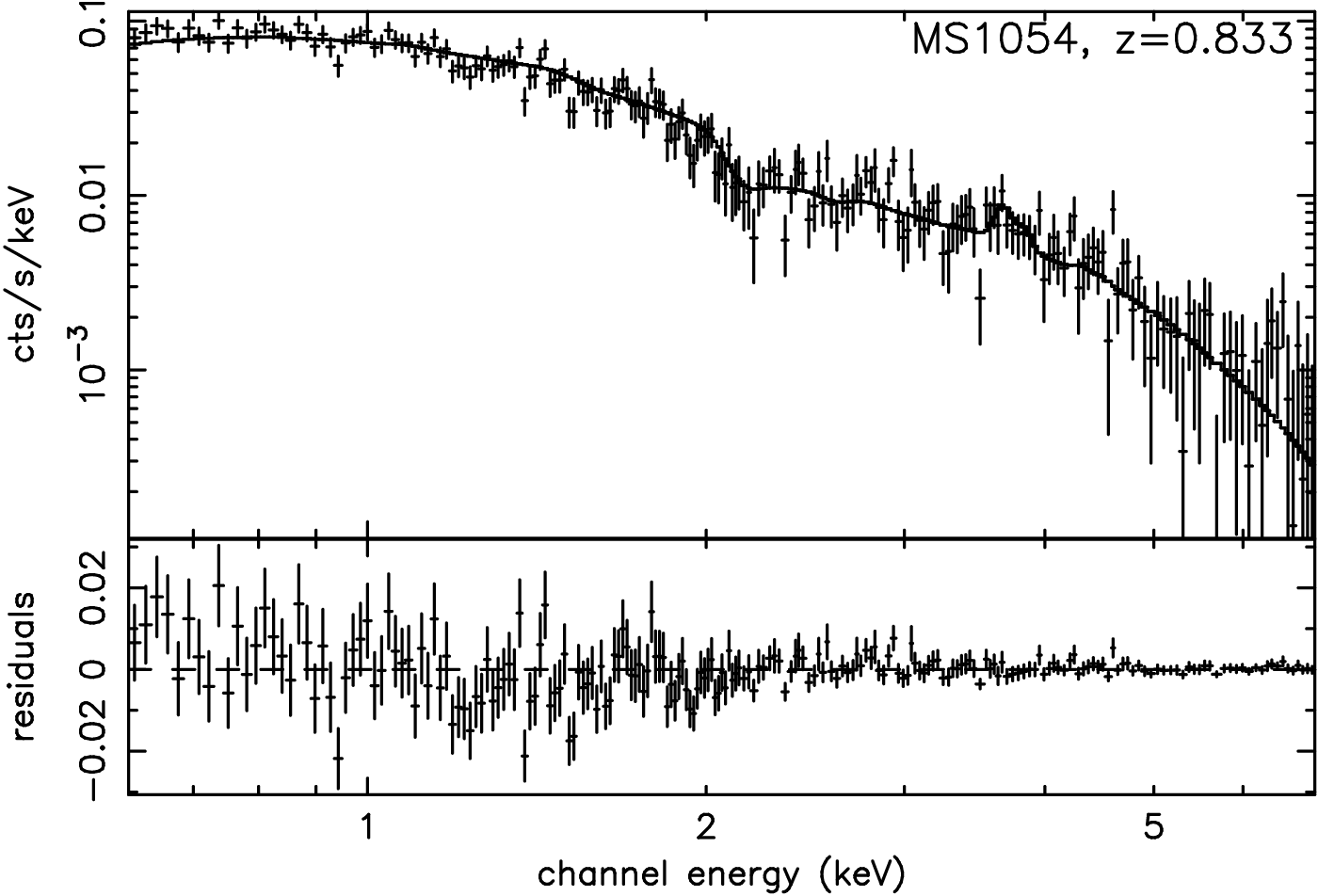,angle=270,height=6cm}}

\caption{The same as Figure \ref{spectra1}.}
\label{spectra2} 
\end{figure}

\begin{figure}
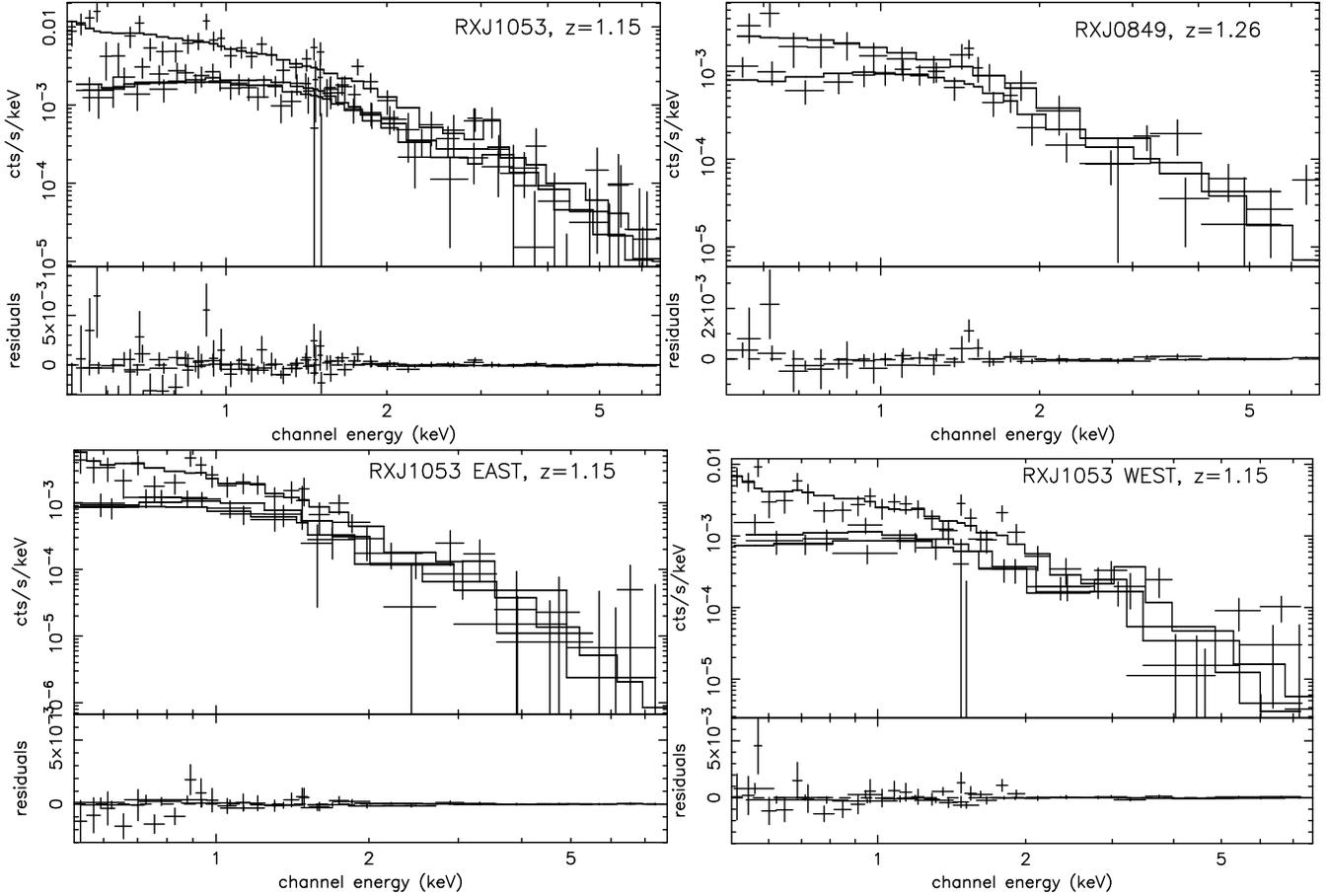
 

\centerline{\psfig{figure=rxj1053_XMM_data.ps,angle=270,height=6cm}\psfig{figure=cl3_XMM_data.ps,angle=270,height=6cm}}

\centerline{\psfig{figure=rxj1053EAST_XMM_data.ps,angle=270,height=6cm}\psfig{figure=rxj1053WEST_XMM_data.ps,angle=270,height=6cm}}

\caption{The spectra of the clusters observed with XMM with the folded
thermal best--fit model (continuous lines).  The lower panels show the
residuals.  The three spectra for RXJ1053 refer to PN, MOS thick
filter and MOS thin filter data while two spectra of RXJ0849 refer to
PN and MOS data.  We also show spectra and best fits for the eastern
and western clumps of RXJ1053.  }
\label{spectra4} 
\end{figure}

\begin{figure}
\centerline{\psfig{figure=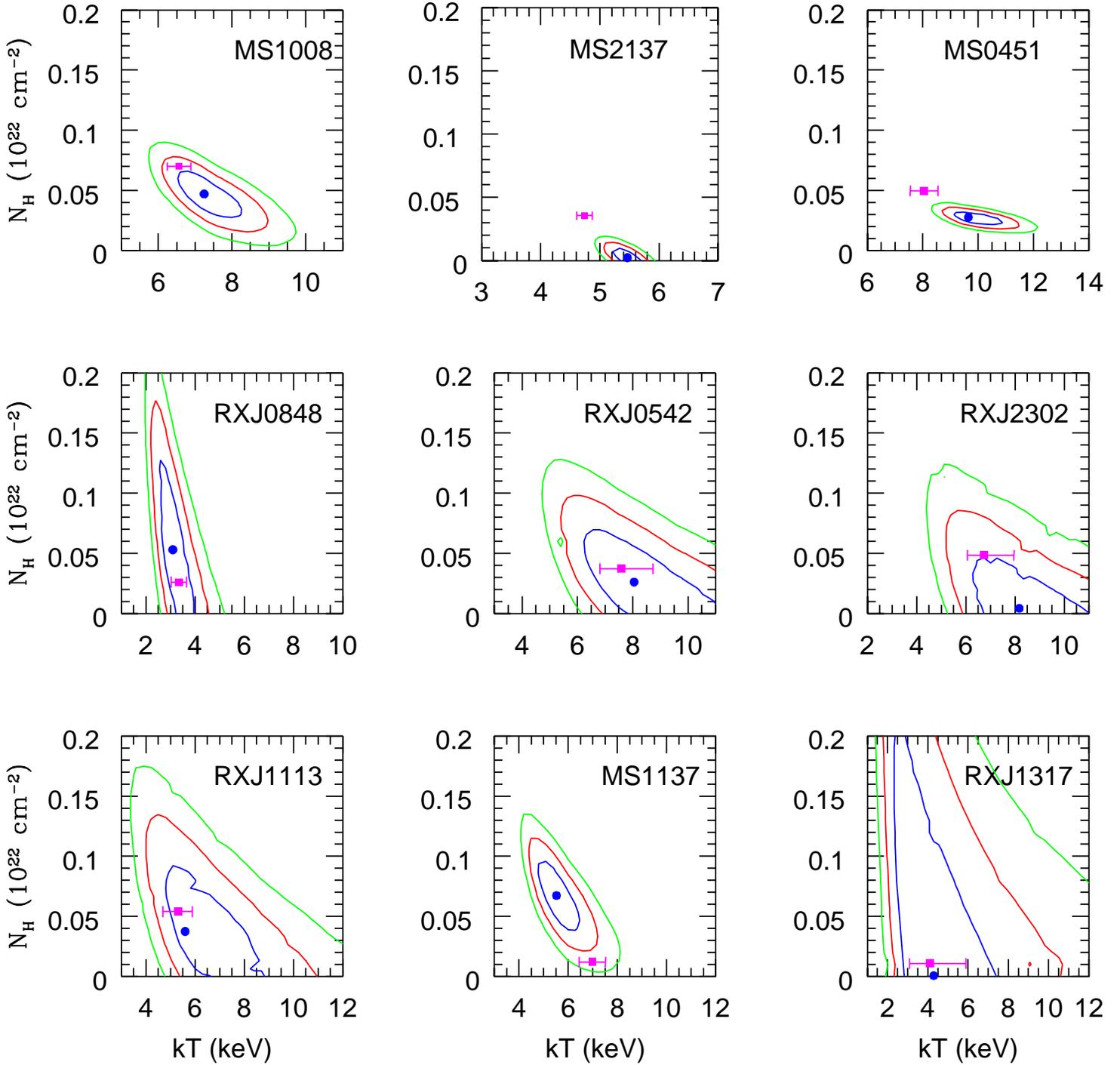,height=8.in,width=8.in}}
\caption{The plots show the 1--2--3 $\sigma$ confidence contours for
the local column density $N_H$ and the temperature $kT$ for the first
nine clusters observed with Chandra.  The filled circle is the
best--fit, and the filled square with $1 \sigma$ error bars is
best--fit point obtained fixing $N_H$ to the Galactic value.
\label{fig_nh1}}
\end{figure}

\begin{figure}
\centerline{\psfig{figure=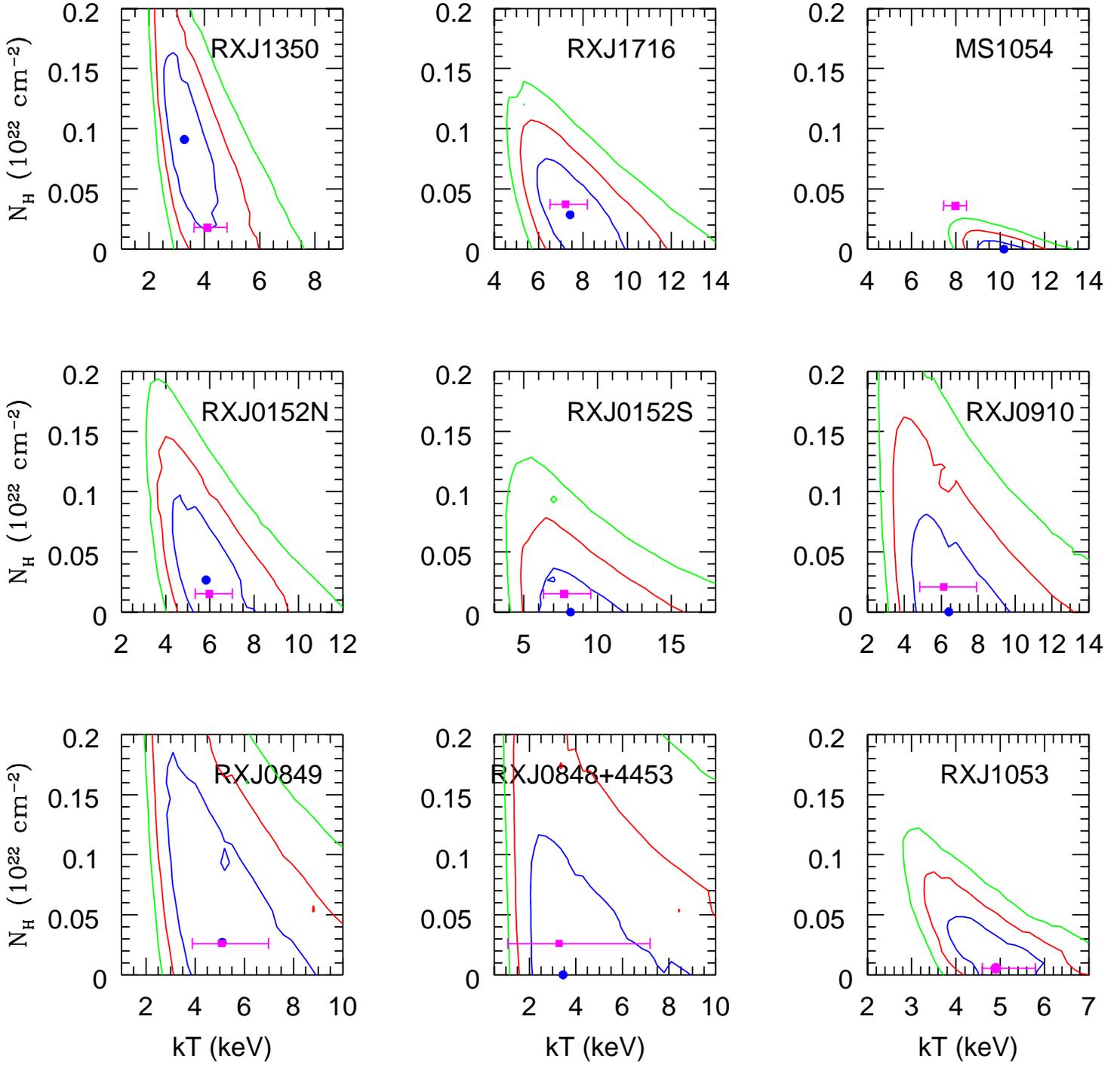,height=8.in,width=8.in}}
\caption{Same as \ref{fig_nh1} for the 8 highest z clusters observed
with Chandra.  The last panel (RXJ1053) refers to XMM only data.  
\label{fig_nh2} }
\end{figure}

In this Appendix, we briefly discuss the spectral analysis performed
for each source of our sample, in order of increasing redshift. In
Figures \ref{spectra1}, \ref{spectra2} and \ref{spectra3}, we show the
background--subtracted spectra, along with the best--fit folded model,
of the clusters observed with Chandra.  In Figure \ref{spectra4} we
show the spectra from the XMM data.  In Figures \ref{fig_nh1} and
\ref{fig_nh2} we also show the $kT$--$N_H$ confidence contours
obtained when the Galactic absorption is left free, and compare the
best-fit values with that obtained freezing $N_H$ to the Galactic
value (see discussion in \S 3).

\subsubsection*{MS1008.1$-$1224} 
This cluster is part of the Extended Medium Sensitivity Survey sample
(EMSS, Gioia et al. 1990). A detailed study of the mass of this system
as derived from Chandra data and weak lensing analysis is presented in
Ettori \& Lombardi (2003). For our spectral analysis we used an
extraction radius of maximum S/N of 108 arc sec, corresponding to 337
$h^{-1}$ kpc for the adopted $\Lambda$CDM cosmology, which includes
$\simeq 9200$ net counts in the 0.3--10 keV band.  Due to the
relatively low redshift ($z=0.306$), emission lines of several
elements are clearly visible in the spectrum around 2 and 5 keV
(observed frame).  The best--fit temperature is $kT=6.57\pm 0.32$ keV,
in agreement with the previous ASCA measurement of
$kT=7.3^{+2.4}_{-1.5}$ keV (Mushotzky \& Scharf 1997) and of
$7.2^{+1.0}_{-0.8}$ keV for the total emission found by Ettori \&
Lombardi (2003).  We do not detect a gradient in the projected
temperature in three annuli out to 100 arc sec.  MS1008 was considered
a prototype of a relaxed cluster, however in a recent optical study
Athreya et al. (2002) found evidence of substructures, possibly
indicating non--equilibrium in the gas, at least in the inner regions.
The best--fit metallicity is $Z=0.24\pm 0.06 Z_\odot$.

\subsubsection*{MS2137.3$-$2353}
 
This very luminous EMSS cluster, at redshift $z=0.313$, was observed
for 43 ks with ACIS--S in VFAINT mode, however the effective exposure
was reduced to 33 ks in our analysis, mostly due to a large flare.
The extraction radius of maximum S/N is 79 arc sec, corresponding to
250$h^{-1}$ kpc.  Inside this circular region we found $\simeq 32000$
net counts in the 0.3--10 keV band.  Due to the high count--rate in
the center of the source, the VFAINT reduction results in the loss of
some source photons.  We repeated the analysis without the VFAINT
cleaning procedure and found that the spectral fit is not affected
(see also {\tt
http://cxc.harvard.edu/cal/Acis/Cal\_prods/vfbkgrnd/index.html\#tth\_sEc4}).
The best--fit temperature is $kT=4.75^{+0.10}_{-0.15}$ keV, in
agreement with the ASCA result, $kT=4.4\pm 0.40$ keV (Mushotzky \&
Scharf 1997).  There is a negative temperature gradient towards the
inner regions.  The temperature in the center is about 3.5 keV,
whereas it is slightly below 6 keV at 30 arc sec, and it decreases
below 4 keV in the outer regions.  Very tight confidence intervals are
obtained for the metallicity: $Z=0.33\pm 0.03 \, Z_\odot$.  This value
is consistent, within 1 $\sigma$, with the ASCA result of $Z=0.41\pm
0.12 \, Z_\odot$ as reported in Mushotzky and Loewenstein (1997).

\subsubsection*{MS0451.6$-$0305} 
This is the most luminous cluster in the EMSS sample. We measured
$\simeq 3400$ net counts from an exposure with ACIS--I (14 ks), and
$\simeq 17000$ net counts from an exposure with ACIS--S (42 ks) in the
0.3--10 keV band, inside an aperture of 98 arc sec radius.  The
best--fit temperature from the combined fit of the two spectra is $kT
= 8.05^{+0.50}_{-0.40}$ keV, in agreement with Vikhlinin et al. (2002)
but lower than the ASCA measure of $kT = 10.2^{+1.5}_{-1.3}$ keV
(Mushotzky \& Scharf 1997).  This difference can be ascribed to a
temperature gradient.  In the Chandra data, the best--fit temperature
ranges from 9 keV in the central 30 arc sec, to 7 keV at radii larger
than 100 arc sec.  The best--fit metallicity is $Z=
0.32^{+0.07}_{-0.04}\, Z_\odot$.  This value is different from the
previous result from ASCA of $Z = 0.15\pm 0.11\, Z_\odot$ (Donahue
1996).  Our analysis of the inner 30 arc sec still gives a quite large
metallicity, $Z=0.60\pm 0.15 Z_\odot$, for a $kT\sim 9$ keV
temperature.  We argue that such a difference can be ascribed to the
details of the analysis of the ASCA data, since a similar discrepance
is found for MS1054 (see below).

\subsubsection*{RXJ0848+4456} 
This cluster, drawn from the ROSAT Deep Cluster Survey (RDCS, Rosati
et al. 1998), is part of a deep observation (184.5 ks of effective
exposure with ACIS--I) of the Lynx field.  Its X--ray and optical
properties and the discovery of a foreground group are discussed in
detail in Holden et al. (2001).  The spectral analysis presents some
difficulties due to a colder, infalling group (RXJ0849+4455) along the
line of sight, at a redshift $z=0.543$, slightly lower than the
redshift of the main cluster, $z=0.570$.  Despite the low number of
net counts detected in the 0.3--10 keV band (850 for the cluster,
within a 30 arc sec circle) the temperature is well constrained since
the cutoff of the thermal spectrum falls just around 2 keV, near the
peak of the efficiency curve of Chandra.  We obtain $kT=3.24\pm 0.30$
keV (in close agreement with Holden et al. (2001) who followed a
slightly different procedure), and a metallicity of $Z
=0.51^{+0.22}_{-0.17}\, Z_\odot$.  We also attempted to derive the
metallicity of the colder group. We have only 200 net counts from a
region of 37.4 arc sec radius. The temperature is well constrained
($kT=2.7^{+1.0}_{-0.6}$ keV), whereas for the metallicity we obtain
only a large upper limit of $Z < 0.6\, Z_\odot$ (1 $\sigma$ upper
limit).  Due to possible contamination from the brighter cluster, we
do not include the colder group in our analysis.

\subsubsection*{RXJ0542.8$-$4100}
This is a cluster from the RDCS sample at $z=0.634$.  We collected
2200 net counts in the 0.3--10 keV band, within an aperture of 79 arc
sec radius.  The best--fit temperature is $kT = 7.6^{+1.1}_{-0.8}$ keV
for a metallicity of $Z = 0.11_{-0.11}^{+0.13}\, Z_\odot$.  We do not
find evidence of a temperature gradient, although the surface
brightness distribution shows a clear envelope, indicative a ``cold
front''.  A detailed study of this cluster will be presented in Scharf
et al.  (in preparation).

\subsubsection*{RXJ2302.8+0844} 
This cluster was discovered in the RDCS at $z=0.734$.  The 110 ks
exposure reduces to an effective time of 108 ks after the standard
reduction.  We detect about 1450 net counts in the 0.3--10 keV band
within a 49 arc sec circle. The best--fit temperature is
$kT=6.7^{+1.3}_{-0.7}$ keV, for a metallicity of
$Z=0.15_{-0.15}^{+0.14}\, Z_\odot$.  The elongated morphology and the
possible evidence for a central double peak suggest that the cluster
may not be in a fully relaxed state.

\subsubsection*{RXJ1113.1$-$2615} 
This cluster was discovered in the WARPS survey (Perlman et al. 2002,
Maughan et al. 2003).  We detected $\sim\! 1200$ net counts in the
0.3--10 keV band from a $39.4$ arc sec radius circular region.  The
high redshift ($z=0.73$), combined with the intermediate temperature,
allowed us to determine the temperature with a 10\% error,
$kT=5.7^{+0.8}_{-0.7}$ keV.  Also the metallicity is well constrained
at a relatively high value, $Z = 0.40\pm 0.18\, Z_\odot$.  Our
results are consistent, within the errors, with the analysis of
Maughan et al. (2003).

\subsubsection*{MS1137.5+6625} 
This is the second most distant cluster in the EMSS survey.  We
detected $\sim 4200$ net counts in the 0.3--10 keV band from a
circular region with a radius of 49 arc sec.  For a redshift of
$z=0.782$, we find a temperature of $kT=7.0\pm 0.5$ keV and a
metallicity of $Z=0.21\pm 0.10 \, Z_\odot$. The best--fit temperature
is consistent with the value of $kT=6.3\pm 0.4$ keV found by Vikhlinin
et al. (2002).  We find a possible drop in the projected temperature
to $\sim 6$ keV in the outer 30--100 arc sec, with low statistical
significance.

\subsubsection*{RXJ1317.4+2911} 
The first analysis of this RDCS cluster was presented by Holden et
al. (2002), along with its optical properties.  The 115 ks exposure
reduced to 110.5 ks after the reduction procedure.  We detected about
230 net counts (in the 0.3--10 keV band) from a circular region with a
radius of 29.5 arc sec, after the removal of pointsources.  The
best--fit temperature is $kT=4.0_{-0.8}^{+1.3}$ keV, for a metallicity
of $Z= 0.52_{-0.38}^{+0.60}\, Z_\odot$, this value is consistent with
the Holden et al. (2002) analysis.  It is, however, higher than that
published by Vikhlinin et al. (2002) of $kT=2.2\pm 0.5$ keV.  The
discrepancy is significant at $2 \sigma$.  We argue that such a
discrepancy can be ascribed to differences in the procedure of
removing faint pointsources within the extraction region, which
becomes critical for clusters with low S/N like this one.  We notice
also that the best--fit redshift from the X--ray analysis alone is
$z\sim 1$, significantly different from the optically measured value
of $z=0.805$.

\subsubsection*{RXJ1350.0+6007} 
The first Chandra analysis of this RDCS cluster, as well its optical
study, was presented by Holden et al. (2002).  We detected about 760
net counts in the 0.3--10 keV band from a circular region of 69 arc
sec radius.  The best--fit temperature is $kT = 4.3_{-0.4}^{+1.3}$
keV, for a metallicity of $0.43_{-0.21}^{+0.44}\, Z_\odot$.  These
values are in close agreement with the Holden et al. (2002) analysis,
and also in very good agreement with Vikhlinin et al. (2002)
($kT=4.3\pm 0.6$ keV).  We found some evidence for a gradient in the
projected temperature, from 3.5 to 5--6 keV going from the central 20
arc sec, to 20--40 and 40--100 arc sec annuli.  A peculiar feature is
that the metallicity is increasing for larger extraction radii
(e.g. the metallicity rises by a factor of two when the redshift is
left free).  This is probably due to noise in the background spectrum,
which becomes larger as the S/N decrease (see the discussion in Holden
et al. 2002).

\subsubsection*{RXJ1716.9+6708} 
This cluster ($z=0.813$) was discovered in the NEP survey, and
presented in Gioia et al. (1999). We use an effective Chandra exposure
of 51 ks.  We detected 1450 net counts in a region of 49.5 arc sec
radius in the 0.3--10 keV band.  The best--fit temperature is $kT =
6.8_{-0.6}^{+1.1} $ keV, for a metallicity of $Z=0.49\pm 0.18 \,
Z_\odot$.  These results are consistent, within 1$\sigma$, with the
values reported by Gioia et al. (1999) using ASCA
($kT=5.7^{+1.4}_{-0.6}$ keV) and by Vikhlinin et al. (2002)
($kT=6.6\pm 0.8$ keV) using Chandra.  We also find some hints of a
higher projected temperature towards the center.

\subsubsection*{MS1054.5$-$0321} 
This is the most distant EMSS cluster, at $z=0.83$, subject to a large
number of follow--up studies.  The 90 ks Chandra exposure reduced to
an effective exposure of 80 ks.  We detected more than 10000 net
counts in the 0.3--10 keV band from an aperture of 78 arc sec radius.
The morphology shows clear evidences of substructures.  We fit the
region of higher S/N without separating the western, cooler clump, as
in Jeltema et al. (2001). We detect a significant temperature gradient
from $\sim 6$ keV in the inner 30 arc sec, to 8 keV in the outer
30--50, 50--80 arc sec rings.  Our best--fit temperature is $kT =
8.0\pm 0.5$ keV, for a metallicity of $0.24^{+0.07}_{-0.08}\,
Z_\odot$.  These results are in good agreement with those by Jeltema
et al. (2001), considering that they analyzed the cooler clump
separately, and by Vikhlinin et al. (2002).  We disagree with the
original ASCA result by Donahue et al. (1998), where the best--fit
temperature was $12.3$ keV and the best--fit metallicity was much
lower.  This situation is similar to that of MS0451, thus indicating
possible problems with the previous analysis of the ASCA data.

\subsubsection*{RXJ0152.7$-$1357} 
This cluster was reported in the RDCS (Della Ceca et al. 2002), in the
WARPS (Ebeling et al. 2000) and SHARC (Romer et al. 2000) surveys.
The system is a typical example of two major mass clumps in the
process of merging, both at redshift $z=0.83$ (Demarco et al., in
preparation). The same morphology is also well traced by the
distribution of cluster galaxies.  An independent detailed analysis of
the Chandra data was recently presented by Maughan et al. (2003).  We
independently analyzed the two clumps which are well separated in the
Chandra image.

For the northern clump we detected $\sim\! 830$ net counts in the
0.3--10 keV band, in a circular region with a radius 58 arc sec.  The
best--fit temperature is $kT=5.9_{-0.7}^{+1.4} $ keV for a metallicity
of $ Z=0.13_{-0.13}^{+0.17}\, Z_\odot $. This result is in excellent
agreement with that derived from global analysis of Beppo--SAX data by
Della Ceca et al. (2000), as well as with that by Vikhlinin et
al. (2002).

For the southern clump we detected $\sim\! 570$ net counts in the
0.3--10 keV band, in a circular region with a radius 53 arc sec.  The
southern clump has a best--fit temperature of $kT=7.6_{-1.0}^{+2.5} $
keV with an upper bound on the metallicity of $Z < 0.19\, Z_\odot$.
This temperature is consistent within $1 \sigma$ with the temperature
of the northern clump.  We notice that if the redshift is left free,
we find for the southern clump a best--fit redshift $z=0.75$ (still
within $1 \sigma$ from the optical redshift $z=0.83$) and a
temperature of $kT\sim 7.4$ keV, but a metallicity jumps to $Z\simeq
0.6 \, Z_\odot$, much higher than that of the northern clump.  Our
temperatures are in good agreement with that of Maughan et al. (2003)
as well as with the analysis of Huo et al. (2003).

\subsubsection*{RXJ0910+5422}  
This is one of the four RDCS clusters at $z>1$ ($z=1.106$). Its
discovery as well as the first analysis of the Chandra data was
presented in Stanford et al. (2002).  It was observed in two pointings
for a total of 173 ks, resulting in an effective exposure of 170 ks.
We detect $\sim\! 430$ net counts in the 0.3--10 keV band from a
region of 25 arc sec. The best--fit temperature is $kT=
6.6_{-1.4}^{+2.0}$ keV.  We obtain only an upper bound for the
metallicity $Z < 0.25 \, Z_\odot$.  These values are in good agreement
with the previous analysis by Stanford et al. (2002).

\subsubsection*{RXJ0849+4452 and RXJ0848+4453} 
These two systems, part of the RDCS sample, are likely part of a
super--structure at $z\simeq 1.26$ (Rosati et al. 1999). They were the
target of a long Chandra exposure (the Lynx field, with an effective
exposure of 184.5 ks).  A detailed analysis of the Chandra
observations was presented in Stanford et al. (2001). At present,
these are the two highest redshift clusters selected and analyzed in
X--rays.

RX0849+4452 shows a relaxed morphology, possibly in an evolved
dynamical stage.  For RX0849+4452 we detect 360 net counts in the
0.3--10 keV band from a circular region of 23.6 arc sec radius in the
Chandra exposure.  The best--fit temperature is $kT =
5.2_{-1.1}^{+1.7}$ keV, with an upper bound on the metallicity of $Z<
0.5\, Z_\odot$.  This result is in excellent agreement with the
previous analysis by Stanford et al. (2002), which gives
$kT=5.8^{+2.8}_{-1.7}$ keV for a fixed metallicity of $Z=0.3\,
Z_\odot$.  We detect also $\sim 630$ net counts from the XMM exposure
(EPIC--PN + MOS detectors).  The analysis of the XMM data for this
cluster yields $kT = 6.6 \pm 1.1$ keV.  We obtain only a $1 \sigma$
upper bound for the metallicity of $Z<0.38 \, Z_\odot$.  These results
are in agreement with the Chandra data within the errors.  The
combined analysis (Chandra + XMM), gives $kT = 5.6_{-0.6}^{+0.8}$ keV,
with a metallicity of $Z = 0.09_{-0.07}^{+0.24} \, Z_\odot$a .  In our
analysis we use the results from the combined fit Chandra + XMM.

As for RXJ0848+4453, we detected $\sim\!130$ net counts in the 0.3--10
keV band in a region of 19.7 arc sec radius.  The best--fit
temperature is $kT= 3.3_{-1.0}^{+2.5}$ keV.  We obtain a very large $1
\sigma$ upper bound for the metallicity $Z < 1.3 \, Z_\odot$.  This
temperature is higher than that derived in the previous
analysis by Stanford et al. (2002) ($kT = 1.6^{+0.8}_{-0.6}$ keV), but
still consistent within less than 2$\sigma$ due to large error bars.
Unfortunately, the XMM data for this cluster are not useful due to the
very low signal--to--noise.  The difference in the best--fit
temperature is in mostly due to the different extraction radius used:
20'' as opposed to the 35'' in Stanford et al. (2001).  By repeating
the analysis with a radius of 35'', we found $kT=2.4_{-0.8}^{+1.2}$
keV, thus in agreement with Stanford et al. (2001).  

\subsubsection*{RXJ1053+5735}  
For this target we used only XMM data from the 100 ks VP observation
of the Lockman Hole (Hashimoto et al. 2002).  The analysis of clearly
shows two clumps.  For the eastern clump we detected about 430 net
counts in the 0.5--8 keV band for the PN camera, and 280 net counts in
the 0.5--8 keV band for the two MOS cameras, in an elliptical region
with semi--axis of 32 and 30.4 arc sec.  The best--fit temperature is
$kT=3.2\pm 0.4 $ keV for a metallicity of $Z=0.60_{-0.22}^{+0.60}\,
Z_\odot $.  For the western clump, we detected about 530 net counts in
the 0.5--8 keV band for the PN camera, and 340 net counts in the
0.5--8 keV band for the two MOS cameras, in an elliptical region with
semi--axis of 27.2 and 28.8 arc sec.  The best--fit temperature is
$kT=5.8_{-0.7}^{+1.2} $ keV for a metallicity of $Z =
0.40^{+0.21}_{-0.14}\, Z_\odot$.  These values are consistent with
those found by Hashimoto et al. (2002), considering that they assumed
$z=1.26$ (Thompson et al. 2001) instead of the revised $z=1.15$
(Hasinger 2003, private communication).

We also present the result of the fit of the total cluster spectrum.
We obtain a best--fit temperature of $kT=4.9_{-0.3}^{+0.9} $ keV for a
metallicity of $Z=0.39_{-0.10}^{+0.14}\, Z_\odot $.  The metallicity
is $Z>0.10 Z_\odot$ at the 90\% c.l., thus constituting the only
detection of the Fe line at a significant confidence level at $z>1$ in
our current sample.

\newpage

\begin{center}
\begin{deluxetable}{l l l l l l l l l }
\footnotesize \tablenum{1} \tablecaption{Chandra Observations
\label{exposures}} \tablewidth{0 pt} \tablehead{Cluster & z & Obs ID &
Exposure & Detectors and obs. mode & $R_{ext}$''& Survey } 
\startdata 
MS1008.1$-$1224 & 0.306 & 926 & 44 & ACIS--I; VFAINT & 108.0 & EMSS &  \nl 
MS2137.3$-$2353 & 0.313 & 928 & 33 & ACIS--S; VFAINT & 79.0 & EMSS  &  \nl 
MS0451.6$-$0305 & 0.539 & 529+902 & 14+42 & ACIS--I/S; VFAINT & 98.0 & EMSS  &  \nl 
RXJ0848+4456 & 0.570 & 1708+927 & 184.5 & ACIS--I; VFAINT  & 30.0 &
RDCS &  \nl
RXJ0542.8$-$4100 & 0.634 & 914 & 50 & ACIS--I; FAINT & 78.7 & RDCS  & \nl 
RXJ2302.8+0844 & 0.734 & 918 & 108 & ACIS--I; FAINT & 49.0 & RDCS  & \nl 
RXJ1113.1$-$2615 & 0.730 & 915 & 103 & ACIS--I; FAINT & 39.4 & WARPS  &  \nl 
MS1137.5+6625  & 0.782 & 536 & 117 & ACIS--I; VFAINT  & 49.2 & EMSS &  \nl 
RXJ1317.4+2911 & 0.805 & 2228 & 110.5 & ACIS--I; VFAINT & 39.4 & RDCS &  \nl
RXJ1350.0+6007 & 0.810 & 2229 & 58 & ACIS--I; VFAINT & 68.9 & RDCS   & \nl
RXJ1716.9+6708 & 0.813 & 548 & 51 & ACIS--I; FAINT & 59.0 & NEP &  \nl
MS1054.5$-$0321  & 0.832 & 512 & 80 & ACIS--S; FAINT  & 78.7 & EMSS  & \nl
RXJ0152.7$-$1357N & 0.835 & 913 & 36 & ACIS--I; FAINT & 58.0 &
RDCS/WARPS/SHARCS  &  \nl
RXJ0152.7$-$1357S & 0.828 & 913 & 36 & ACIS--I; FAINT  & 52.7  &
RDCS/WARPS/SHARCS  &   \nl
RXJ0910+5422 & 1.106 & 2452+2227 & 170 & ACIS--I; VFAINT & 24.6 & RDCS
&  \nl  
RXJ0849+4452 & 1.261 & 1708+927 & 184.5 & ACIS--I; VFAINT & 29.5 & RDCS
&  \nl 
RXJ0848+4453 & 1.273 & 1708+927 & 184.5 & ACIS--I; VFAINT & 29.5  &
RDCS  & \nl
\enddata 
\tablecomments{The sample of 17 clusters observed with Chandra.  
The 3rd columns is the ID of the exposure in the Chandra Archive.  The
4th column refers to the effective exposure time in ks, after removal
of high background intervals.  The extraction radius
$R_{ext}$ is given in arc seconds.}
\end{deluxetable}
\end{center}

\begin{center}
\begin{deluxetable}{l l l l l ll}
\footnotesize \tablenum{2} \tablecaption{XMM Observations
\label{exposuresxmm}} \tablewidth{0 pt} \tablehead{Cluster & z
& exposure & detector & $R_{ext}$'' & survey } 
\startdata 
RXJ1053+5735 & 1.15\footnote{Revised redshift, Hasinger private communication} &  94.5 & PN + 2 MOS & 32  & Lockman Hole\nl 
RXJ0849+4452 & 1.26 &  112.0 & PN + 2 MOS & 29.5  & RDCS\nl 
\enddata 
\tablecomments{The two XMM observations considered in this Paper.     The
3rd column refers to the effective exposure time in ks, after removal
of high background intervals.  The 4th column shows the detectors
used.  The extraction radius $R_{ext}$ is given in arc seconds.}
\end{deluxetable}
\end{center}

\newpage

\begin{center}
\begin{deluxetable}{l l l l l }
\footnotesize \tablenum{3} \tablecaption{Spectral fits from Chandra
and XMM observations
\label{results}} \tablewidth{0 pt} 
\tablehead{Cluster & kT (keV) & $Z/Z_\odot$ & $N_H$ (cm$^{-2}$)}
\startdata MS1008.1$-$1224 & $6.57\pm 0.32$ & $0.24\pm 0.06$ &
$7.26\times 10^{20}$ \nl 
MS2137.3$-$2353 & $4.75^{+0.10}_{-0.15}$ & $0.33 \pm 0.03$ &
$3.55\times 10^{20}$ 
\nl MS0451.6$-$0305 & $8.05^{+0.50}_{-0.40}$ & $0.32^{+0.07}_{-0.04}$
& $5.00\times
10^{20}$ \nl 
RXJ0848+4456 & $3.24\pm 0.30$ & $0.51^{+0.22}_{-0.17}$ & $2.63\times
10^{20}$  \nl
RXJ0542.8$-$4100 & $7.6^{+1.1}_{-0.8}$ & $0.11_{-0.11}^{+0.13}$ &
$3.73\times 10^{20}$  \nl 
RXJ2302.8+0844 & $6.7_{-0.7}^{+1.3}$ & $0.15_{-0.15}^{+0.14}$ &
$4.85\times 10^{20}$ \nl 
RXJ1113.1$-$2615 & $5.7^{+0.8}_{-0.7}$ & $0.40\pm 0.18$ & $5.50\times
10^{20}$ \nl
MS1137.5+6625 & $7.0\pm 0.5 $ & $0.21^{+0.10}_{-0.11} $ & $1.21\times
10^{20}$ \nl 
RXJ1317.4+2911 & $4.0_{-0.8}^{+1.3}$ & $0.52_{-0.38}^{+0.60}$ &
$1.10\times 10^{20}$ \nl 
RXJ1350.0+6007 & $4.3_{-0.4}^{+1.3}$ & $0.43_{-0.21}^{+0.44}$ &
$1.80\times 10^{20}$ \nl 
RXJ1716.9+6708 & $6.8_{-0.6}^{+1.1} $ & $0.49\pm 0.18 $ &
$ 3.72\times 10^{20}$ \nl 
MS1054.5$-$0321 & $8.0 \pm 0.5$ & $0.24_{-0.08}^{+0.07}$ & $3.61\times
10^{20}$ \nl 
RXJ0152.7$-$1357N & $5.9_{-0.7}^{+1.4} $ & $ 0.13_{-0.13}^{+0.17}$ &
$1.54\times 10^{20}$ \nl 
RXJ0152.7$-$1357S & $7.6_{-1.0}^{+2.5} $ & $ <0.19$ & $1.54 \times
10^{20}$ \nl 
RXJ0910+5422 & $6.6_{-1.4}^{+2.0}$ & $<0.25$ & $ 2.10\times 10^{20}$\nl 
RXJ0849+4452 & $5.6_{-0.6}^{+0.8}$ & $0.09^{+0.24}_{-0.07}$ &
$2.63\times 10^{20}$ \nl 
RXJ0848+4453 & $3.3_{-1.0}^{+2.5}$ & $0.25_{-0.25}^{+0.95}$ &
$2.63\times 10^{20}$ \nl 
RXJ1053+5735 &$4.9_{-0.3}^{+0.9}$ & $0.39_{-0.10}^{+0.14}$ & $5.7\times
10^{19}$ \nl
RXJ1053+5735 EAST &$3.2\pm 0.4$ & $0.60^{+0.60}_{-0.22}$ &
$5.7\times 10^{19}$  \nl 
RXJ1053+5735 WEST &$5.8_{-0.7}^{+1.2}$ & $0.40_{-0.14}^{+0.21}$ & $5.7\times
10^{19}$ 
\nl \enddata

\tablecomments{Results from the spectral fits with the
TBABS$\times$MEKAL model.  The local column density is always frozen
at the galactic value, measured by Dickey \& Lockman (1990) and shown
in the 4th column.  Error bars refer to $1 \sigma$ confidence level.
}
\end{deluxetable}
\end{center}

\newpage

\begin{center}
\begin{deluxetable}{l l l l }
\footnotesize \tablenum{4} \tablecaption{Average Iron Abundance from
combined fits
\label{tableZ}} \tablewidth{0 pt} 
\tablehead{ $\langle z\rangle $ & $Z/Z_\odot$ ($kT> 5$ keV) &$Z/Z_\odot$ ($kT< 5$ keV) }
\startdata 
0.30 &  $0.30^{+0.03}_{-0.2}$ [2]&  -- \nl
0.58 &  $0.29^{+0.07}_{-0.03}$ [2]   &  $0.51^{+0.22}_{-0.17}$ [1]  \nl
0.80 &  $0.25^{+0.04}_{-0.06}$  [7] &  $0.50^{+0.29}_{-0.22}$  [2] \nl
1.20  &  $0.22^{+0.12}_{-0.10}$ [3] &  $0.65^{+0.50}_{-0.30}$ [2]  \nl 
\enddata
\tablecomments{The first comlumn shows the average redshift of the bin.
The number of the clusters of each combined fit is shown in
parenthesis.  }
\end{deluxetable}
\end{center}

\newpage 
 
\newpage

\newpage
 
\newpage 

\newpage
 
\newpage
 
\newpage

\newpage
 
\newpage
 
\begin{figure} 

\centerline{\psfig{figure=cl_913_up_data.ps,angle=270,height=6cm}\psfig{figure=cl_913_down_data.ps,angle=270,height=6cm}}

\centerline{\psfig{figure=cl_2227_data.ps,angle=270,height=6cm}\psfig{figure=cl_cl3_data.ps,angle=270,height=6cm}}

\centerline{\psfig{figure=cl_cl1_data.ps,angle=270,height=6cm}}

\caption{The same as Figure \ref{spectra1}.}
\label{spectra3} 
\end{figure}

\newpage
 
\newpage

\newpage
  
\end{document}